\documentclass[10pt]{iopart}
\usepackage{graphicx}
\usepackage{tabls}
\usepackage{bm}
\eqnobysec

\newcommand{\beq}{\begin{equation}}
\newcommand{\beqa}{\begin{eqnarray}}
\newcommand{\eeq}{\end{equation}}
\newcommand{\eeqa}{\end{eqnarray}}

\newcommand{\abs}[1]{\vert#1\vert}
\newcommand{\bigabs}[1]{\left\vert#1\right\vert}
\newcommand{\mean}[1]{\langle#1\rangle}
\newcommand{\bigmean}[1]{\left\langle#1\right\rangle}
\newcommand{\dd}{{\rm d}}
\newcommand{\eps}{\varepsilon}
\newcommand{\erf}{{\mathop{\rm erf}}}
\newcommand\frad[2]{{\displaystyle{#1\over #2}}}

\newcommand{\ii}{{\rm i}}
\renewcommand{\max}{{\rm max}}
\renewcommand{\min}{{\rm min}}
\newcommand{\hrho}{{\widehat\rho}}
\newcommand{\pyro}{{\rm (pyro)}}

\newcommand{\wt}{\widetilde}
\newcommand{\1}{{(1)}}
\newcommand{\2}{{(2)}}
\newcommand{\Ai}{{\mathop{\rm Ai}}}
\newcommand{\Bi}{{\mathop{\rm Bi}}}
\newcommand{\C}{{\cal C}}
\newcommand{\E}{{\cal E}}
\newcommand{\I}{{\rm I}}
\newcommand{\II}{{\rm II}}
\renewcommand{\L}{{\rm (L)}}
\renewcommand{\Re}{\mathop{\rm Re}\,}
\renewcommand{\Im}{\mathop{\rm Im}\,}

\def\Xint#1{\mathchoice
{\XXint\displaystyle\textstyle{#1}}
{\XXint\textstyle\scriptstyle{#1}}
{\XXint\scriptstyle\scriptscriptstyle{#1}}
{\XXint\scriptscriptstyle\scriptscriptstyle{#1}}
\!\int}
\def\XXint#1#2#3{{\setbox0=\hbox{$#1{#2#3}{\int}$}
\vcenter{\hbox{$#2#3$}}\kern-.5\wd0}}
\def\dashint{\Xint-}

\begin{document}

\title{Scaling laws for weakly disordered 1D flat bands}

\author{J M Luck}

\address{Institut de Physique Th\'eorique, Universit\'e Paris-Saclay, CEA and CNRS,
91191~Gif-sur-Yvette, France}

\begin{abstract}
We investigate Anderson localization on various 1D structures having flat bands.
The main focus is on the scaling laws obeyed by the localization length
at weak disorder in the vicinity of flat-band energies.
A careful distinction is made between situations where the scaling functions are universal
(i.e., depend on the disorder distribution only through its width)
and where they keep depending on the full shape of the disorder distribution,
even in the weak-disorder scaling regime.
Three examples are analyzed in detail.
On the stub chain,
one central flat band is isolated from two lateral dispersive ones.
The localization length remains microscopic at weak disorder
and exhibits disorder-specific features.
On the pyrochlore ladder,
the two flat bands are tangent to a dispersive one.
The localization length diverges with exponent 1/2
and a non-universal scaling law,
whose dependence on the disorder distribution is predicted analytically.
On the diamond chain,
a central flat band intersects two symmetric dispersive ones.
The localization length exhibits two successive scaling regimes,
diverging first with exponent 4/3 and a universal law,
and then (i.e., further away from the pristine flat band)
with exponent 1 and a non-universal law.
Both scaling functions are also derived by analytical means.
\end{abstract}

\ead{\mailto{jean-marc.luck@ipht.fr}}

\maketitle

\section{Introduction}
\label{intro}

Tight-binding electronic spectra of crystalline structures having more than one atom per unit cell
may exhibit flat bands.
A flat band is a dispersionless branch of the spectrum,
such that energy $E$ is independent of wavevector $\bm q$.
Equivalently,
the tight-binding Hamiltonian has an extensively degenerate set of eigenstates
at that value of energy,
with a basis of eigenstates that are strictly localized,
i.e., supported by finitely many sites.
Such an extensive (i.e., macroscopic) degeneracy induces an extremely high sensitivity
to various kinds of perturbations, including interactions, disorder, or external fields.
A weak disordered potential provides an example of foremost importance of such a perturbation.
Any amount of disorder is indeed sufficient, in generic circumstances,
to induce enough hybridization to entirely lift the extensive degeneracy.
Flat-band systems have actually been shown to exhibit Anderson localization
with various kinds of unconventional features~\cite{god,nis,cha,nit,S1,S2,GLN,BM}.
More specific predictions have been made in the one-dimensional situation,
where eigenstate are exponentially localized,
irrespective of the disorder width,
at least in generic circumstances~\cite{gang,lgp,fifty}.
The key question therefore resides in the scaling behavior of the localization length $\xi$
as a function of the disorder width $w$ in the weak-disorder regime.
For a usual dispersive band, $\xi$ is known to diverge with exponent 2 inside the band,
and with exponent $2/3$ in the vicinity of band edges.
More exotic exponents have been recently reported
for various weakly disordered 1D structures having flat bands,
including 0 for the stub chain,
$1/2$ for the cross-stitch and pyrochlore ladders,
and $4/3$ for the diamond chain and Lieb ladder~\cite{FL1,FL2,FL3,FL4,Ge}.

The goal of this paper is to corroborate the above predictions
on weakly disordered~1D flat-band systems
and, more importantly, to complement them with quantitative results
on the corresponding scaling laws (i.e., full scaling functions)
in the weak-disorder regime.
The main focus will be on whether these laws
are universal (i.e., depend on the distribution of disorder only through its width $w$)
or depend on the full shape of the disorder distribution.
We shall study three examples in detail.
In all three cases, the analysis will be reduced
to investigating the Lyapunov exponent of products of random $2\times2$ transfer matrices
in the weak-disorder regime.
The key object of the approach is the invariant distribution of the Riccati variables
(section~\ref{secchain}).
The stub chain (section~\ref{secstub})
has an isolated flat band and scaling exponent~0, meaning
that the localization length remains microscopic at weak disorder.
Its non-universal features are only accessible by perturbative and numerical means.
On the pyrochlore ladder (section~\ref{secpyro}),
the two flat bands are tangent to a dispersive one.
The localization length has scaling exponent $1/2$.
The associated non-universal law is predicted analytically.
On the diamond chain (section~\ref{secdiam}),
a central flat band intersects two symmetric dispersive ones.
This is the richest of all examples.
The localization length exhibits two successive scaling regimes,
diverging first with exponent $4/3$ and a universal law,
and then (i.e., further away from the pristine flat band)
with exponent 1 and a non-universal law.
Both scaling functions are predicted by analytical means.
Section~\ref{disc} summarizes our findings.
Two appendices contain more technical material.
Cauchy tails of probability distributions are investigated in~\ref{appcau},
whereas~\ref{appeff} is devoted to the effective disorder
on the pyrochlore ladder and the diamond chain.

\section{A reminder on localization on the chain}
\label{secchain}

This section is a reminder on localization properties
of the tight-binding model on the chain with a weak diagonal disorder.
This gives us the opportunity of introducing concepts and tools
which will be at work in the sequel.

The tight-binding equation on the chain reads
\beq
E\psi_n=\psi_{n+1}+\psi_{n-1}+v_n\psi_n.
\label{tb}
\eeq
The random potentials $v_n$ are independent from each other
and drawn from some symmetric (i.e., even) distribution $\rho(v)$.

On a clean chain (i.e., without disorder),
the spectrum consists of a single band with dispersion relation
\beq
E=2\cos q\qquad(\abs{q}\le\pi).
\eeq

\subsection{Transfer-matrix approach and Lyapunov exponent}

The transfer-matrix approach consists in recasting~(\ref{tb}) as
\beq
\pmatrix{\psi_{n+1}\cr\psi_n}=T_n\pmatrix{\psi_n\cr\psi_{n-1}},
\eeq
where the transfer matrix
\beq
T_n=\pmatrix{E-v_n&-1\cr 1&0}
\label{tn}
\eeq
obeys $\det T_n=1$.
We have thus
\beq
\pmatrix{\psi_{n+1}\cr\psi_n}=T_n\dots T_1\pmatrix{\psi_1\cr\psi_0}.
\eeq
The exponential growth (or decay) of a generic wavefunction is therefore
dictated by the exponential growth of a large product of random matrices.
In particular, the Lyapunov exponent of the matrix product,
\beq
\gamma=\lim_{n\to\infty}\frac{1}{n}\ln\bigabs{T_n\dots T_1},
\eeq
identifies with the inverse localization length~\cite{fur,bor,car,pzis,alea,ctt}.

In the following, we recall the scaling properties of the Lyapunov exponent,
both inside the band and near the band edges.
The cases of regular disorder with a finite variance
and of Cauchy disorder will be dealt with successively.

\subsection{Regular disorder}

Regular disorder corresponds to the case where the disorder variance $\mean{v^2}=w^2$
is finite (i.e., convergent).
The quantity $w$ is referred to as the disorder width.
The scaling properties of the Lyapunov exponent at weak disorder are universal,
in the sense that the disorder distribution only enters through~$w$.

Inside the band, the Lyapunov exponent scales as
\beq
\gamma\approx\frac{w^2}{8\sin^2q}\approx\frac{w^2}{2(4-E^2)}.
\label{regband}
\eeq
This result, known as the Thouless formula~\cite{thh},
breaks down near band edges ($E\to\pm2$),
indicating that eigenstates are more strongly localized there than inside the band.
The Lyapunov exponent obeys the scaling law~\cite{flo,hal,dgedge,irt,cltt}
\beq
\gamma\approx(w^2/2)^{1/3}\;F\!\left(\frac{\abs{E}-2}{(w^2/2)^{2/3}}\right),
\label{regedge}
\eeq
where the universal scaling function
\beq
F(x)=\frac{\Ai(x)\Ai'(x)+\Bi(x)\Bi'(x)}{\Ai(x)^2+\Bi(x)^2}
\eeq
involves the Airy functions $\Ai$ and $\Bi$.

\subsection{Cauchy disorder (the Lloyd model)}
\label{Lloyd}

The Cauchy distribution
\beq
\rho(v)=\frac{W}{\pi(v^2+W^2)}
\eeq
is a very special instance of a disorder distribution with an infinite variance.
First, the tight-binding model with Cauchy disorder,
known as the Lloyd model~\cite{lloyd},
is solvable on any regular lattice in any dimension,
in the sense that the average one-particle Green's function can be evaluated exactly.
Second, an effective Cauchy disorder emerges in several instances
of 1D structures possessing flat bands~\cite{FL1,FL2,FL3,FL4,Ge},
including the pyrochlore ladder and the diamond chain,
investigated in sections~\ref{secpyro} and~\ref{secdiam}.

Let us give a full derivation of the Lyapunov exponent for Cauchy disorder on the chain,
following the approach introduced by Dyson~\cite{Dy} and Schmidt~\cite{Sch}.
The key quantities are the Riccati variables, defined as the ratios
\beq
R_n=\frac{\psi_{n+1}}{\psi_n}.
\eeq
They obey the recursion
\beq
R_n=E-v_n-\frac{1}{R_{n-1}},
\label{rec}
\eeq
so that $R_n$ depends on $R_0$ and on the random potentials $v_1,\dots,v_n$.
Its probability density $f_n(R_n)$ obeys the integral recursion formula
\beqa
f_n(R_n)&=&\int_{-\infty}^\infty\rho(v_n)\dd v_n
\int_{-\infty}^\infty f_{n-1}(R_{n-1})\dd R_{n-1}
\nonumber\\
&\times&\delta\left(R_n-E+v_n+\frac{1}{R_{n-1}}\right)
\nonumber\\
&=&\int_{-\infty}^\infty\frac{\rho(v_n)\dd v_n}{(E-v_n-R_n)^2}\,f_{n-1}\left(\frac{1}{E-v_n-R_n}\right).
\label{DS}
\eeqa
When $n$ becomes infinitely large,
$f_n(R_n)$ converges to a limiting distribution~$f(R)$,
referred to as the invariant distribution associated with the random recursion~(\ref{rec}).
The Lyapunov exponent reads
\beq
\gamma=\Re\mean{\ln R}=\mean{\ln\abs{R}},
\label{gammares}
\eeq
where the average is taken w.r.t.~the invariant distribution.

For Cauchy disorder,~(\ref{DS}) can be worked out explicitly.
Assuming $R_{n-1}$ has a Cauchy distribution of the form
\beq
f(R_{n-1})=\frac{b_{n-1}}{\pi((R_{n-1}-a_{n-1})^2+b_{n-1}^2)},
\eeq
the recursion~(\ref{DS}) implies that $R_n$ is also Cauchy distributed,
with parameters
\beq
a_n=E-\frac{a_{n-1}}{a_{n-1}^2+b_{n-1}^2},\qquad
b_n=W+\frac{b_{n-1}}{a_{n-1}^2+b_{n-1}^2}.
\eeq
The above recursion assumes a simpler form in terms of the complex variables
\beq
z_n=a_n+\ii b_n,\qquad\E=E+\ii W,
\eeq
namely
\beq
z_n=\E-\frac{1}{z_{n-1}}.
\eeq
The complex variables $z_n$ converge to the stable fixed point
of this transformation,
\beq
z_\star=a_\star+\ii b_\star=\frac{\E+\sqrt{\E^2-4}}{2}.
\eeq
The invariant distribution $f(R)$ is therefore a Cauchy distribution,
whose parame\-ters~$a_\star$ and $b_\star$ are the real and imaginary parts of $z_\star$.
Equation~(\ref{gammares}) yields
\beq
\gamma=\frac{\ln(a_\star^2+b_\star^2)}{2}=\ln\abs{z_\star}.
\eeq

More explicit predictions can be made at weak-disorder ($W\ll1$).
Inside the band, the Lyapunov exponent scales as
\beq
\gamma\approx\frac{W}{2\sin q}.
\label{Cauband}
\eeq
The above estimate again breaks down near the band edges.
Consider the upper band edge for definiteness.
Introducing the scaling variable
\beq
x=\frac{E-2}{W},
\eeq
we have
\beq
z_\star\approx1+\sqrt{(x+\ii)W}.
\label{zstar}
\eeq
The Lyapunov exponent therefore obeys the scaling law
\beq
\gamma\approx W^{1/2}\,F^\L(x),
\label{Cauedge}
\eeq
whose scaling function reads
\beq
F^\L(x)=\Re\sqrt{x+\ii}=\frac{1}{\sqrt{2}}\left(x+\sqrt{x^2+1}\right)^{1/2}.
\eeq

Anderson localization on the chain thus exhibits different
scaling laws for a weak regular disorder and a weak Cauchy disorder.
For regular disorder,
the universal formulas~(\ref{regband}) within the band
and~(\ref{regedge}) near band edges
have respective exponents~2 and 2/3.
For Cauchy disorder,
the formulas~(\ref{Cauband}) within the band and~(\ref{Cauedge}) near band edges
have respective exponents 1 and 1/2.
It is also worth noting that the universal scaling function $F(x)$
entering~(\ref{regedge}) in the regular case has a much richer structure
than $F^\L(x)$ entering~(\ref{Cauedge}) in the Cauchy case.

\section{The stub chain}
\label{secstub}

\begin{figure}[!ht]
\begin{center}
\includegraphics[angle=0,width=.45\linewidth]{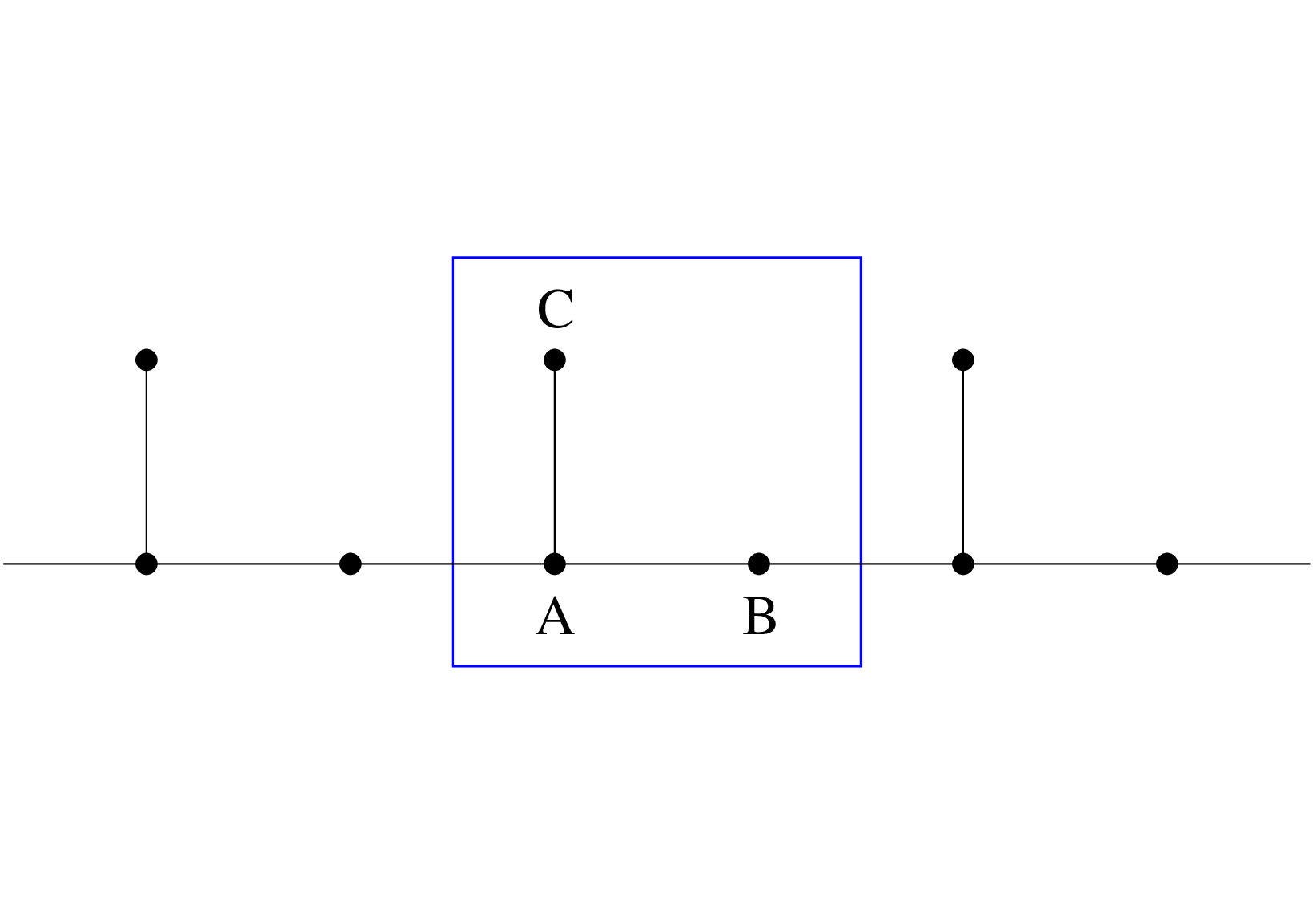}
\hskip 10pt
\includegraphics[angle=0,width=.35\linewidth]{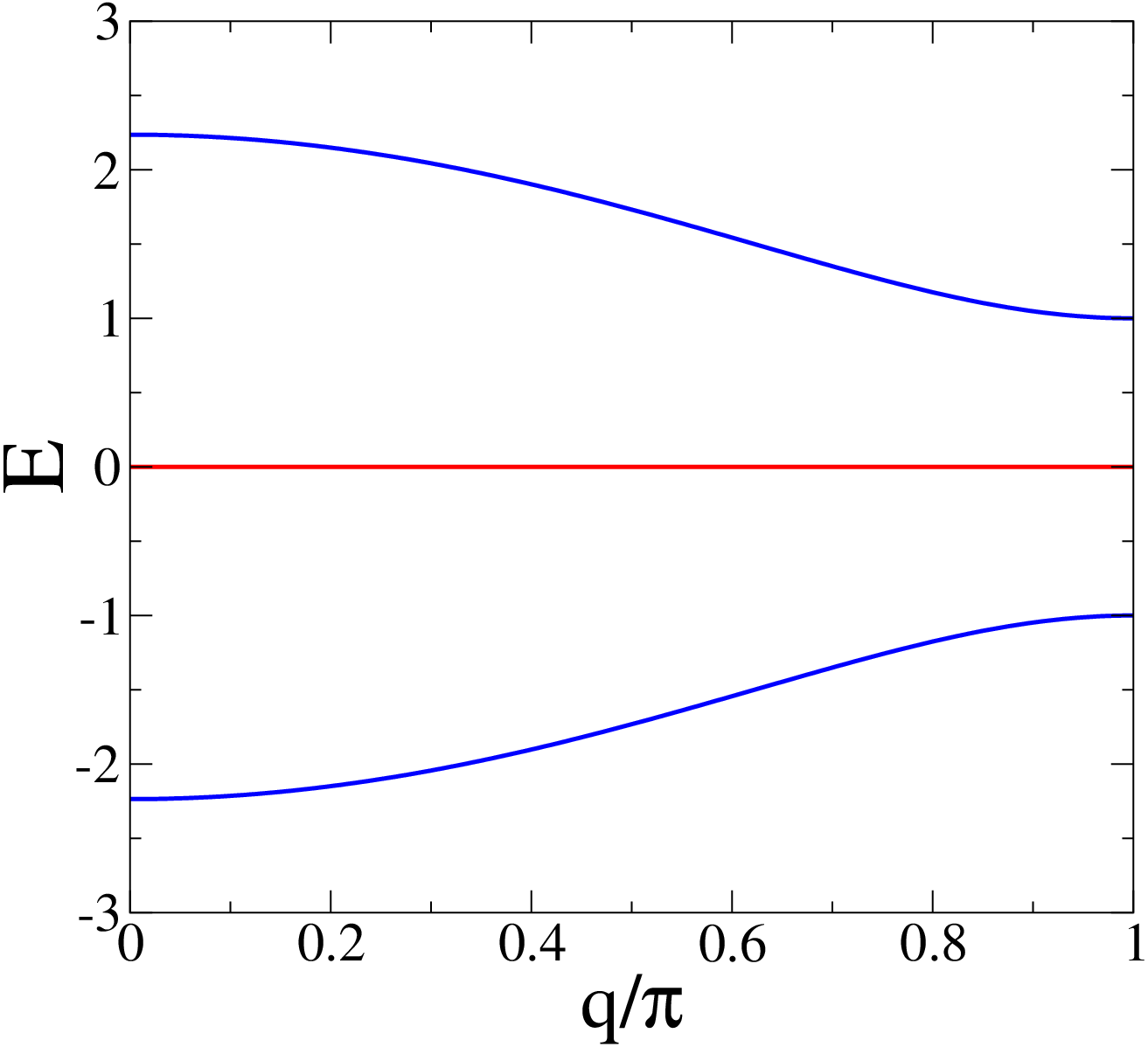}
\caption{\small
Left: the stub chain has 3 atoms per unit cell.
Right: its spectrum (see~(\ref{spstub})) has one central flat band (red)
isolated from two symmetric dispersive ones (blue).}
\label{stub}
\end{center}
\end{figure}

The stub chain (see figure~\ref{stub}, left)
is the simplest of all 1D structures exhibiting a flat band~\cite{FL2,FL3,Ge}.
We equip it with a tight-binding model with diagonal disorder.
All hopping amplitudes between neighboring atoms equal unity.
All disordered potentials $v_n^A$, $v_n^B$, $v_n^C$
are drawn from the same symmetric distribution $\rho(v)$ with variance $\mean{v^2}=w^2$.
The tight-binding equations at energy $E$ read
\beqa
(E-v_n^A)\psi_n^A&=&\psi_{n-1}^B+\psi_n^B+\psi_n^C,
\nonumber\\
(E-v_n^B)\psi_n^B&=&\psi_n^A+\psi_{n+1}^A,
\nonumber\\
(E-v_n^C)\psi_n^C&=&\psi_n^A.
\label{tbstub}
\eeqa

The dispersion relation of a clean stub chain
is obtained by expressing that $E$ is an eigenvalue of the matrix
\beq
M(q)=\pmatrix{0&1+\e^{-\ii q}&1\cr 1+\e^{\ii q}&0&0\cr 1&0&0}
\eeq
encoding the right-hand sides of~(\ref{tbstub}).
The spectrum (see figure~\ref{stub}, right)
consists of one flat band and two dispersive bands:
\beqa
E&=&0,
\nonumber\\
E&=&\pm\sqrt{3+2\cos q}.
\label{spstub}
\eeqa
The flat band is isolated, i.e., separated from the dispersive ones
by a finite gap $\Delta E=1$.
It consists of the extensively degenerate states
\beq
\psi_n^B=-\sigma_n,\qquad\psi_n^C=\sigma_{n-1}+\sigma_n,
\label{degenestub}
\eeq
where $\sigma_n$ are arbitrary amplitudes.
The most localized states, corresponding e.g.~to $\sigma_n=\delta_{n0}$,
live on three sites belonging to two consecutive cells, such as e.g.~$B_0$, $C_0$ and $C_1$.

In the presence of a weak disorder,
the flat band is broadened symmetrically over an energy range of order $w$.
If the disorder distribution is supported by a finite interval,
i.e., $\rho(v)$ is non-zero only for $\abs{v}<V$,
the disordered flat band has well-defined edges at $E=\pm V$.
The density of states vanishes exponentially fast in the vicinity of these band edges,
which correspond to internal Lifshitz tails~\cite{klopp1,klopp2}.
For $V<1/2$,
the disordered flat band remains isolated from the dispersive ones.
For the uniform distribution on $[-V,V]$, with $w^2=V^2/3$,
the band edges read $E=\pm w\sqrt{3}$.

The weakly disordered flat band of the stub lattice has been
investigated in several recent works~\cite{FL2,FL3,Ge}.
The subsequent analysis corroborates earlier findings
and allows us to make more precise statements on some aspects.

Whenever energy $E$ and all disordered potentials are simultaneously small,
the tight-binding equations~(\ref{tbstub}) simplify to
\beqa
\psi_{n+1}^A&\approx&-\psi_n^A+\eta_n\phi_n^B,
\nonumber\\
\phi_{n+1}^B&\approx&-\phi_n^B-\chi_n\psi_{n+1}^A,
\label{tbnew}
\eeqa
where the amplitudes $\psi_n^C$ have been eliminated,
the rescaling $\phi_n^B=E\psi_n^B$ has been done, and
\beq
\eta_n=1-\frac{v_n^B}{E},\qquad
\chi_n=\left(1-\frac{v_n^C}{E}\right)^{-1}
\eeq
are reduced disorder variables.
Furthermore, the rescaled equations~(\ref{tbnew})
can be recast as a product of $2\times2$ independent random matrices, namely
\beq
\pmatrix{\psi_{n+1}^A\cr\phi_{n+1}^B}\approx T_n\pmatrix{\psi_n^A\cr\phi_n^B},
\eeq
with
\beq
T_n=-\pmatrix{1&-\eta_n\cr-\chi_n&1+\eta_n\chi_n}.
\eeq
We have again $\det T_n=1$.
The associated Riccati variables obey the recursion
\beq
R_{n+1}=\frac{R_n-\eta_n}{1+\chi_n(\eta_n-R_n)}.
\label{rstub}
\eeq
The associated Lyapunov exponent
\beq
\gamma=\bigmean{\ln\bigabs{\frac{R_n-\eta_n}{R_n}}}
\label{gstub}
\eeq
gives the inverse localization length, in units of the size of a unit cell.

The above formalism involves $E$ only through the dimensionless quantities
$\eta_n$ and~$\chi_n$.
As a consequence, the Lyapunov exponent
depends on energy only through the dimensionless ratio~\cite{FL3,Ge}
\beq
x=\frac{E}{w}.
\label{epsdef}
\eeq

In the absence of disorder ($\eta_n=\chi_n=1$),
the recursion~(\ref{rstub}) has two fixed points,~$R_+$ (unstable)
and $R_-$ (stable), with $R_\pm=(1\pm\sqrt{5})/2$.
The associated Lyapunov exponent is $\gamma_\infty=\ln(1-1/R_-)$, i.e.,
\beq
\gamma_\infty=\ln\frac{3+\sqrt{5}}{2}=0.962423.
\label{ginf}
\eeq
This is the value of the inverse localization length
for $w\ll\abs{E}\ll1$, i.e., $\abs{x}\gg1$.
This result can be recovered by more elementary means,
along the lines of~\cite{Ge}.
The pristine chain has extensively degenerate states given by~(\ref{degenestub}).
An infinitesimal disorder lifts this extensive degeneracy
by allowing the flat-band states to hybridize with the dispersive ones.
Setting $E=0$ in the expression~(\ref{spstub}) of the dispersive bands
yields a complex momentum $q_\star$ such that $\cos q_\star=-3/2$,
i.e., $\e^{\ii q_\star}=-(3\pm\sqrt{5})/2$.
The latter momentum describes the falloff of flat-band states
in the limit of an infinitesimal disorder.
The identification $\abs{\Im q_\star}=\gamma_\infty$ allows one
to recover~(\ref{ginf}).

A systematic weak-disorder expansion around~(\ref{ginf})
can be derived by considering the variables
\beq
Z_n=\frac{R_n-R_-}{R_n-R_+},
\eeq
which represent the deviation of the Riccati variables from the stable fixed point,
and are therefore small at weak disorder.
The $Z_n$ obey the recursion
\beq
Z_{n+1}=\frac{N_{n+1}}{D_{n+1}},
\label{zrec}
\eeq
with
\beqa
N_{n+1}&=&\left(\sqrt{5}-5+(1-\sqrt{5})v_n^B+4v_n^C+(1+\sqrt{5})v_n^Bv_n^C\right)\!(Z_n-1)
\nonumber\\
&+&\sqrt{5}-5+(5+\sqrt{5})v_n^C,
\nonumber\\
D_{n+1}&=&\left(-(4+2\sqrt{5})v_n^B-(1+\sqrt{5})v_n^C+(1+\sqrt{5})v_n^Bv_n^C\right)\!(Z_n-1)
\nonumber\\
&-&4\sqrt{5}-10+(5+\sqrt{5})v_n^C,
\eeqa
whereas~(\ref{gstub}) reads
\beq
\gamma=\bigmean{\ln\bigabs{1+(\sqrt{5}+1)\frac{(1-v_n^B)(1-Z_n)}{2+(3+\sqrt{5})Z_n}}}.
\eeq
Expanding the recursion~(\ref{zrec}) in powers of the disordered potentials
allows one to derive a weak-disorder expansion of the moments
of the invariant distribution of $Z$.
We thus obtain
\beq
\mean{Z}=-\frac{5+\sqrt{5}}{50}\frac{\mean{v^2}}{E^2}
+\frac{29\sqrt{5}-75}{250}\frac{\mean{v^4}}{E^4}
+\frac{27-12\sqrt{5}}{125}\frac{\mean{v^2}^2}{E^4}+\cdots
\eeq
and analogous expressions for $\mean{Z^p}$ for $p=2$, 3 and 4.
Putting everything together,
we obtain a fourth-order expansion of the Lyapunov exponent.
Once recast in terms of the reduced quantities $x$ (see~(\ref{epsdef}))
and $\kappa=\mean{v^4}/w^4$, the kurtosis of the disorder distribution,
the latter expansion appears as a large-$x$ expansion:
\beq
\gamma=\gamma_\infty
+\frac{\sqrt{5}-1}{5\,x^2}+\frac{(6\sqrt{5}-8)\kappa+5-3\sqrt{5}}{25\,x^4}+\cdots
\label{gammapert}
\eeq
Higher-order dimensionless ratios such as $\kappa_p=\mean{v^{2p}}/w^{2p}$
for $p=3,4,\dots$~enter higher-order terms of the above expansion.
This demonstrates that the Lyapunov exponent depends on all moments
of the disorder distribution, i.e., on its full shape,
and not only on the disorder width $w$.

\begin{figure}[!ht]
\begin{center}
\includegraphics[angle=0,width=.5\linewidth]{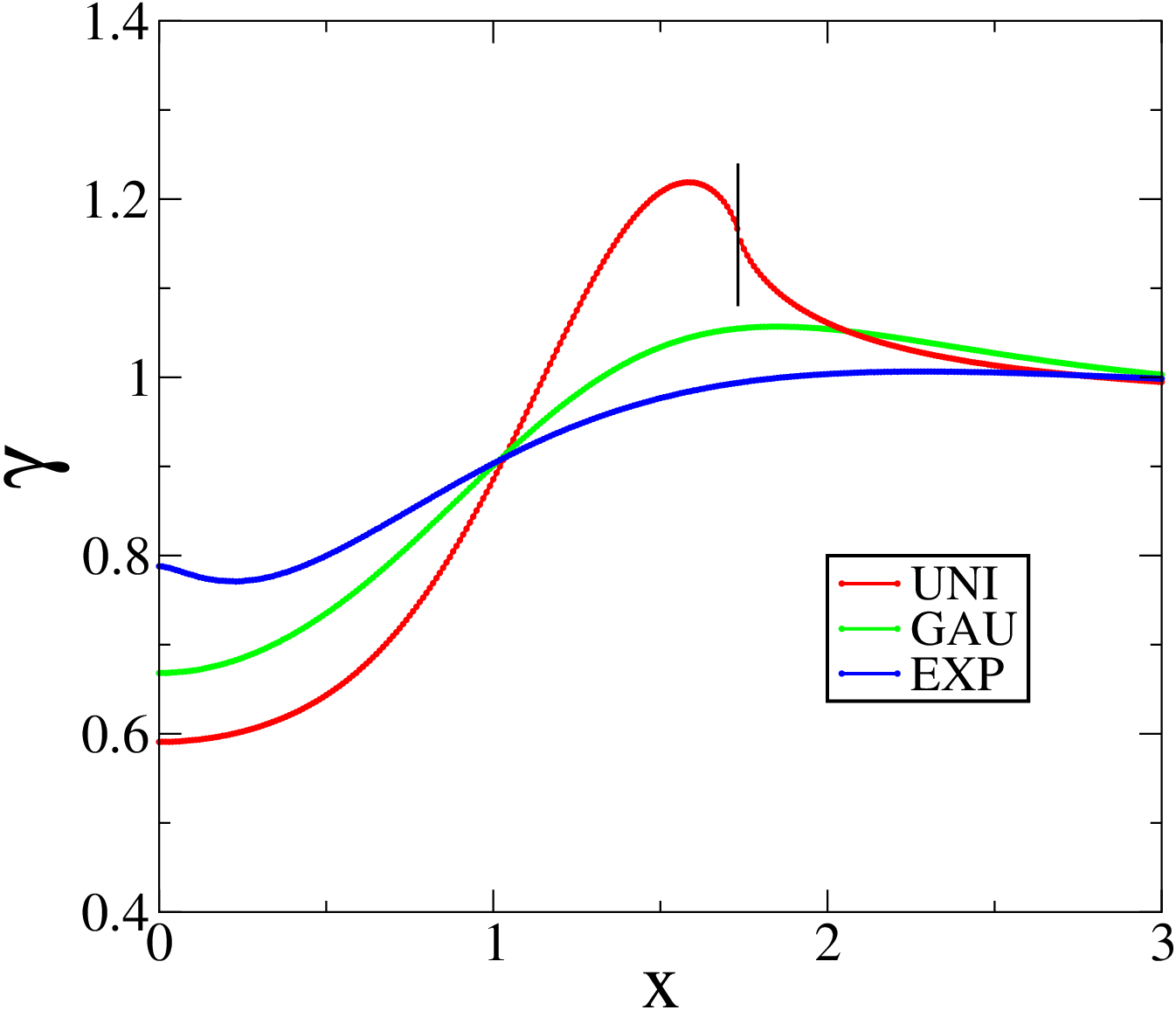}
\caption{\small
The Lyapunov exponent $\gamma$ against reduced energy $x=E/w$
for three disorder distributions:
uniform, Gaussian and exponential.
Vertical segment: edge of flat band for the uniform distribution ($x=\sqrt3$).}
\label{gammastub}
\end{center}
\end{figure}

Besides the above expansion, no analytical result is available in general.
In order to proceed, we must have recourse to a numerical study.
A very efficient approach consists in building up the invariant measure
by iterating the rescaled recursion~(\ref{rstub}) along a very long trajectory
(e.g.~of length $10^8$).
Figure~\ref{gammastub} shows the Lyapunov exponent thus obtained.
Throughout this work, we compare results for three disorder distributions:
uniform, Gaussian and exponential (see~\ref{appdis} for details).

The Lyapunov exponent (inverse localization length) is an even smooth function
of the reduced energy $x$ (see~(\ref{epsdef})),
which depends on the shape of the disorder distribution.
It exhibits a non-monotonic dependence on the reduced energy,
with a non-trivial maximum $\gamma_\max$,
put forward in Reference~\cite{Ge} as an anomalous minimum localization length.
For large $x$, it approaches the limit $\gamma_\infty$ from above,
according to the expansion~(\ref{gammapert}),
which also predicts the observed ordering of the curves at large~$x$.
The coefficient of the kurtosis $\kappa$ in~(\ref{gammapert}) is positive,
and the curves are indeed ordered as the kurtosis of the distributions,
which are recalled in table~\ref{tabstub} below.
For the uniform distribution, the disordered flat band
exhibits a band edge at $x=\sqrt{3}$, shown as a vertical segment.
The Lyapunov exponent reaches its maximum within the band
but not far from the band edge.
It seems to have only a very mild singularity there,
in agreement with the Lifshitz behavior of the density of states recalled above.
The Lyapunov exponent often reaches its minimum at the band center
($\gamma_\min=\gamma(0)$),
in agreement with the intuitive picture that the states closest to the band center
are the most hybridized, so that their localization length is the largest.
This is however not true for the exponential distribution,
where the Lyapunov exponent exhibits a non-trivial minimum $\gamma_\min$ slightly below $\gamma(0)$.
All the characteristic values mentioned above
depend on the shape of the disorder distribution (see table~\ref{tabstub}).
The numbers for uniform and Gaussian disorders agree with those to be read off
from the figures of~\cite{Ge}.
Finally, for the distributions we have considered,
all these quantities are monotonic functions of the kurtosis $\kappa$.

\begin{table}[!ht]
\begin{center}
\begin{tabular}{|l|c|c|c|c|}
\hline
Distribution & $\kappa$ & $\gamma_\max$ & $\gamma(0)$ & $\gamma_\min$\\
\hline
Uniform & 1.8 & 1.219 & 0.591 & --\\
Gaussian & 3 & 1.057 & 0.668 & --\\
Exponential & 6 & 1.006 & 0.788 & 0.771\\
\hline
\end{tabular}
\caption{Characteristic values of the Lyapunov exponent (inverse localization length)
of the weakly disordered flat band of the stub lattice
for three disorder distributions,
whose kurtosis $\kappa$ is recalled
(see~\ref{appdis} for details).
The minimum $\gamma_\min$ is given when it is non-trivial,
i.e., different from $\gamma(0)$.}
\label{tabstub}
\end{center}
\end{table}

\section{The pyrochlore ladder}
\label{secpyro}

\begin{figure}[!ht]
\begin{center}
\includegraphics[angle=0,width=.45\linewidth]{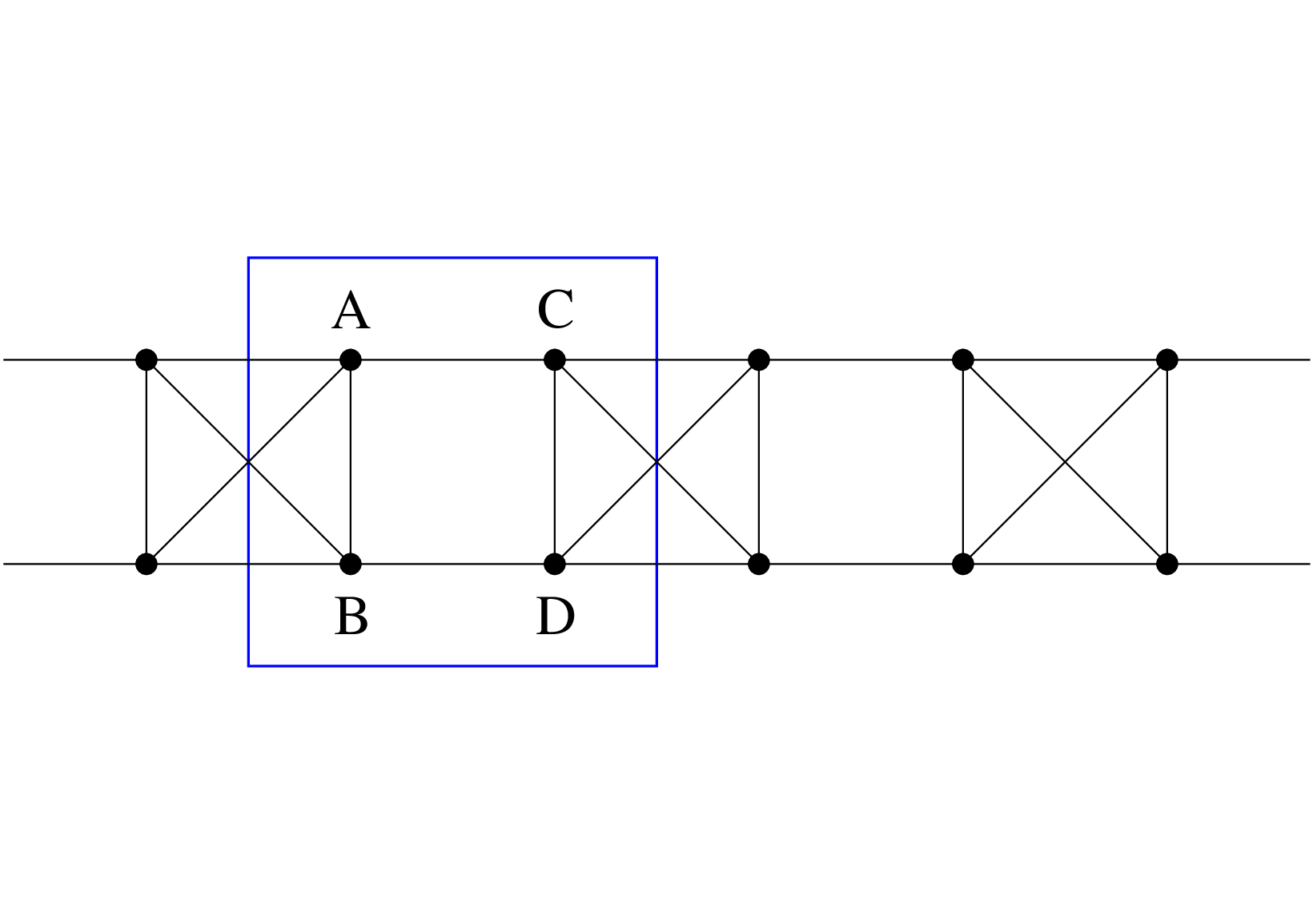}
\hskip 10pt
\includegraphics[angle=0,width=.35\linewidth]{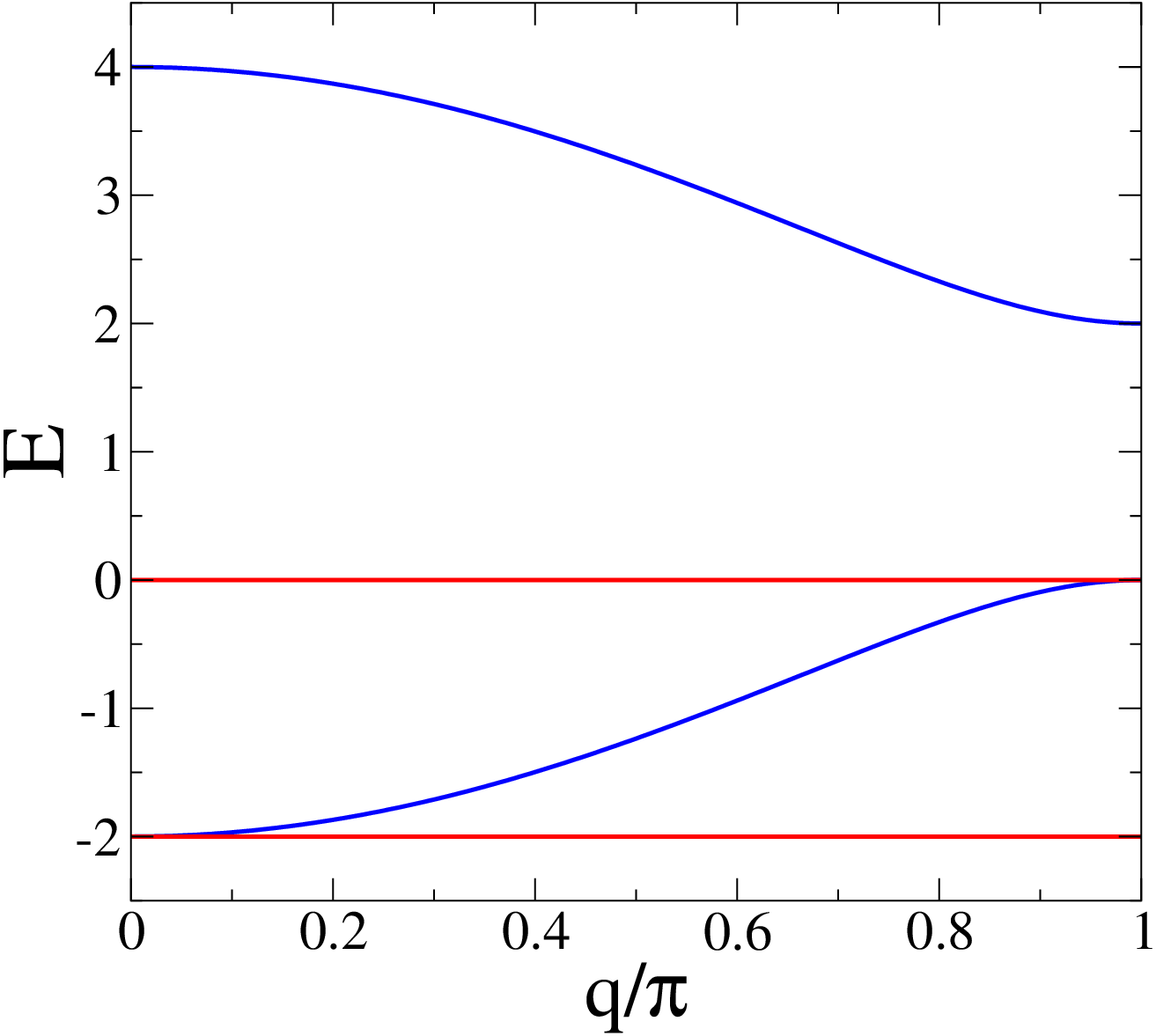}
\caption{\small
Left: the pyrochlore ladder has 4 atoms per unit cell.
Right: its spectrum (see~(\ref{sppyro})) has two flat bands (red)
and two dispersive ones (blue).}
\label{pyro}
\end{center}
\end{figure}

The pyrochlore ladder (see figure~\ref{pyro}, left)
is a well-known example of a 1D structure exhibiting flat bands~\cite{FL2,FL3}.
We equip it with a tight-binding model with diagonal disorder.
All disordered potentials are again drawn from the same distribution.
The tight-binding equations at energy $E$ read
\beqa
(E-v_n^A)\psi_n^A&=&\psi_{n-1}^C+\psi_{n-1}^D+\psi_n^B+\psi_n^C,
\nonumber\\
(E-v_n^B)\psi_n^B&=&\psi_{n-1}^C+\psi_{n-1}^D+\psi_n^A+\psi_n^D,
\nonumber\\
(E-v_n^C)\psi_n^C&=&\psi_n^A+\psi_n^D+\psi_{n+1}^A+\psi_{n+1}^B,
\nonumber\\
(E-v_n^D)\psi_n^D&=&\psi_n^B+\psi_n^C+\psi_{n+1}^A+\psi_{n+1}^B.
\label{tbpyro}
\eeqa

The dispersion relation of a clean ladder
is obtained by expressing that $E$ is an eigenvalue of the matrix
\beq
M(q)=\pmatrix{0&1&1+\e^{-\ii q}&\e^{-\ii q}\cr 1&0&\e^{-\ii q}&1+\e^{-\ii q}\cr
1+\e^{\ii q}&\e^{\ii q}&0&1\cr\e^{\ii q}&1+\e^{\ii q}&1&0}.
\eeq
The spectrum (see figure~\ref{pyro}, right)
consists of two flat bands and two dispersive ones:
\beqa
E&=&0,
\nonumber\\
E&=&-2,
\nonumber\\
E&=&1\pm\sqrt{5+4\cos q}.
\label{sppyro}
\eeqa
The lower dispersive band touches both flat bands.
The latter consist of the extensively degenerate states
\beqa
E=0:\hskip 32.5pt\psi_n^A=\psi_n^C=\sigma_n,\quad\psi_n^B=\psi_n^D=-\sigma_n,
\nonumber\\
E=-2:\qquad\psi_n^A=\psi_n^D=\tau_n,\quad\psi_n^B=\psi_n^C=-\tau_n,
\label{degenepyro}
\eeqa
where $\sigma_n$ and $\tau_n$ are arbitrary amplitudes.
The most localized states,
corresponding e.g.~to $\sigma_n=\delta_{n0}$ or $\tau_n=\delta_{n0}$,
live on the four sites of a single unit cell.

The weakly disordered flat bands of the pyrochlore ladder
have been investigated recently~\cite{FL2,FL3}.
Here, at variance with the case of the stub chain, a dispersive band touches both flat bands,
so that the localization length is expected to be large when disorder is weak.
From a quantitative viewpoint,
a scaling of the localization length in $w^{-1/2}$ is predicted in References~\cite{FL2,FL3},
based on the occurrence of an effective Cauchy disorder.
The subsequent analysis corroborates this finding,
provides an exact mapping onto the Lloyd model at weak disorder
and yields the analytical prediction~(\ref{fpyro})
for the corresponding non-universal scaling function $F^\pyro(x)$.

Let us focus our attention onto the flat band at $E=0$ and introduce the combinations
\beqa
s_n&=&\psi_n^A+\psi_n^B+\psi_n^C+\psi_n^D,
\nonumber\\
t_n&=&\psi_n^A+\psi_n^B-\psi_n^C-\psi_n^D,
\nonumber\\
u_n&=&\psi_n^A-\psi_n^B+\psi_n^C-\psi_n^D,
\nonumber\\
v_n&=&\psi_n^A-\psi_n^B-\psi_n^C+\psi_n^D.
\eeqa
In the absence of disorder,
the tight-binding equations~(\ref{tbpyro}) imply that
$s_n$ and $t_n$ obey the recursion
\beqa
s_{n+1}=t_n-s_n,
\nonumber\\
t_{n+1}=-t_n,
\label{st}
\eeqa
while $v_n$ vanishes and $u_n$ is left arbitrary.
The latter property is expected, as $u_n=4\sigma_n$ describes
the compactly localized states on the pristine ladder (see~(\ref{degenepyro})).

The idea of the subsequent analysis is to follow how the recursion~(\ref{st}),
involving two variables only, is deformed
when energy~$E$ and all random potentials are simultaneously small.
A careful treatment of the tight-binding equations~(\ref{tbpyro})
to first order in the variables
\beq
\eta_n^A=E-v_n^A,\quad
\eta_n^B=E-v_n^B,\quad
\eta_n^C=E-v_n^C,\quad
\eta_n^D=E-v_n^D
\eeq
yields
\beq
\pmatrix{s_{n+1}\cr t_{n+1}}\approx M_n\pmatrix{s_n\cr t_n},
\eeq
with
\beqa
M_n
&=&\pmatrix{-1-\alpha_{n+1}&1+\alpha_{n+1}-\beta_{n+1}\cr
\gamma_n+\alpha_{n+1}&-1-\delta_n+\beta_{n+1}-\alpha_{n+1}}
\nonumber\\
&\approx&\pmatrix{-1-\alpha_{n+1}&1+\alpha_{n+1}-\beta_{n+1}\cr
\alpha_{n+1}&-1+\beta_{n+1}-\alpha_{n+1}}
\pmatrix{1-\gamma_n&\delta_n\cr -\gamma_n&1+\delta_n}
\label{sdef}
\eeqa
and
\beqa
\alpha_n=\frac{1}{\Delta_n}(2\eta_n^A\eta_n^B+\eta_n^A\eta_n^D+\eta_n^B\eta_n^C),
\nonumber\\
\beta_n=\frac{1}{\Delta_n}(2\eta_n^A\eta_n^B+\eta_n^A\eta_n^C+\eta_n^B\eta_n^D),
\nonumber\\
\gamma_n=\frac{1}{\Delta_n}(2\eta_n^C\eta_n^D+\eta_n^A\eta_n^D+\eta_n^B\eta_n^C),
\nonumber\\
\delta_n=\frac{1}{\Delta_n}(2\eta_n^C\eta_n^D+\eta_n^A\eta_n^C+\eta_n^B\eta_n^D),
\eeqa
with
\beq
\Delta_n=2(\eta_n^A+\eta_n^B+\eta_n^C+\eta_n^D).
\eeq
We have $\alpha_n+\delta_n=\beta_n+\gamma_n$.

In the weak-disorder regime,
the Lyapunov exponent of the above matrix product can be determined
by means of an exact mapping onto the Lloyd model (see section~\ref{Lloyd}).
The mapping proceeds in two steps.
First, the transfer matrices $M_n$ are not independent from each other,
as they involve the disordered potentials in two neighboring cells.
This can be remedied by swapping both matrices entering the second line
of~(\ref{sdef}), i.e., by replacing $M_n$ by
\beqa
\wt M_n&=&-\pmatrix{1-\gamma_n&\delta_n\cr -\gamma_n&1+\delta_n}
\pmatrix{-1-\alpha_n&1+\alpha_n-\beta_n\cr\alpha_n&-1+\beta_n-\alpha_n}
\nonumber\\
&\approx&
\pmatrix{1+a_n&-1+c_n\cr b_n&1+d_n},
\label{wmnpyro}
\eeqa
to leading order,
with
\beq
a_n=\alpha_n-\gamma_n,\quad
b_n=-\alpha_n-\gamma_n,\quad
c_n=2\delta_n,\quad
d_n=2\gamma_n.
\eeq
We have $a_n+b_n+d_n=0$, and so $\det\wt M_n=1$, to leading order.
A minus sign has been introduced for convenience in the definition of $\wt M_n$.
Second, a change of basis defined by the matrix
\beq
P=\pmatrix{1&0\cr -1&1}
\eeq
amounts to replacing $\wt M_n$ by
\beq
T_n=P^{-1}\wt M_nP\approx\pmatrix{2+a_n-c_n&-1+c_n\cr 1-c_n-2d_n&c_n+d_n}.
\label{tnpyro}
\eeq
At weak disorder, the above transfer matrix is close to that of the Anderson model
on the chain, given by~(\ref{tn}), near the upper band edge ($E\approx2$).
The Riccati variables associated with~(\ref{tnpyro}) obey the recursion
\beq
R_n=\frac{(2+a_n-c_n)R_{n-1}-1+c_n}{(1-c_n-2d_n)R_{n-1}+c_n+d_n}.
\label{recpyro}
\eeq
Let us anticipate for a while that the disorder entering~(\ref{tnpyro}) has Cauchy tails.
An appropriate identification of the random recursions~(\ref{rec}) and~(\ref{recpyro})
yields a mapping of the present problem onto the Lloyd model
(see section~\ref{Lloyd}).
Let us denote quantities pertaining to the latter model with a superscript $\L$.
Near the upper band edge ($E^\L\approx2$) and at weak disorder,
the invariant distribution $f^\L(R)$ is a narrow distribution concentrated around $R=1$.
Equation~(\ref{zstar}) indeed shows that $\mean{R}=a_\star^\L$ is close to unity,
whereas the width $b_\star^\L$ scales as $\sqrt{W^\L}$.
It is therefore sufficient to compare the formulas~(\ref{rec}) and~(\ref{recpyro})
in the vicinity of $R=1$.
Both equations obey $\partial R_n/\partial R_{n-1}\approx1$ at weak disorder.
Equating the values of $R_n$ for $R_{n-1}=1$ yields the identification
\beq
E^\L-2-v_n^\L\approx
a_n+d_n=\alpha_n+\gamma_n=\frac{\eta_n^\1\eta_n^\2}{\eta_n^\1+\eta_n^\2},
\label{iden}
\eeq
to leading order,
with
\beqa
\eta_n^\1=\eta_n^A+\eta_n^C=2E-v_n^A-v_n^C,
\nonumber\\
\eta_n^\2=\eta_n^B+\eta_n^D=2E-v_n^B-v_n^D.
\eeqa

The effective disorder defined in the rightmost side of~(\ref{iden})
is a homogeneous but non-linear combination of the four disordered potentials of the $n$th unit cell.
It is typically small at weak disorder.
It may however become large in the rare instances where its denominator nearly vanishes.
As a consequence,
it has a broad probability distribution exhibiting symmetric Cauchy tails (see~\ref{appcau}).
In the present context,
this specific feature of disorder has been put forward for the first time in Reference~\cite{FL1}.
The above effective disorder is therefore characterized by two parameters,
its Cauchy strength and its mean value, namely
\beq
G_2(E)=\bigmean{\delta\left(\frac{\eta_n^\1+\eta_n^\2}{\eta_n^\1\eta_n^\2}\right)},
\qquad
H_2(E)=\bigmean{\frac{\eta_n^\1\eta_n^\2}{\eta_n^\1+\eta_n^\2}}.
\label{g2h2}
\eeq
These quantities are investigated in~\ref{apppyro}.
Their scaling dependence on $w$ reads (see~(\ref{g2h2app}))
\beq
G_2(E)=wg_2(x),\qquad H_2(E)=wh_2(x),\qquad x=\frac{E}{w}.
\eeq
Working out explicitly the identification~(\ref{iden}) yields
\beq
W^\L\approx\pi wg_2(x),\qquad x^\L\approx\frac{h_2(x)}{\pi g_2(x)}.
\label{map}
\eeq

We finally obtain our prediction for the scaling behavior
of the Lyapunov exponent of the weakly disordered flat band near $E=0$
by inserting the estimates~(\ref{map}) into~(\ref{Cauedge}).
This reads
\beq
\gamma\approx w^{1/2}\,F^\pyro(x),
\label{pyrores}
\eeq
where the scaling function reads
\beq
F^\pyro(x)=
\frac{1}{\sqrt{2}}\left(h_2(x)+\sqrt{h_2(x)^2+\pi^2g_2(x)^2}\right)^{1/2}.
\label{fpyro}
\eeq
This result simplifies right at $E=0$.
We have $h_2(0)=0$ by symmetry, and so
\beq
F^\pyro(0)=\left(\frac{\pi g_2(0)}{2}\right)^{1/2}.
\eeq

The scaling function $F^\pyro(x)$ is non-universal,
as it depends on the full shape of the disorder distribution
through the functions $g_2(x)$ and $h_2(x)$,
investigated in~\ref{apppyro}.
For the uniform, Gaussian and exponential distributions,
explicit expressions of~$g_2(x)$ are given in~\ref{appdis},
and the values of $g_2(0)$ in table~\ref{tabdis}.
The function~$h_2(x)$ is expressed as an integral
that can be carried out in closed form only for Gaussian disorder.

\begin{figure}[!ht]
\begin{center}
\includegraphics[angle=0,width=.49\linewidth]{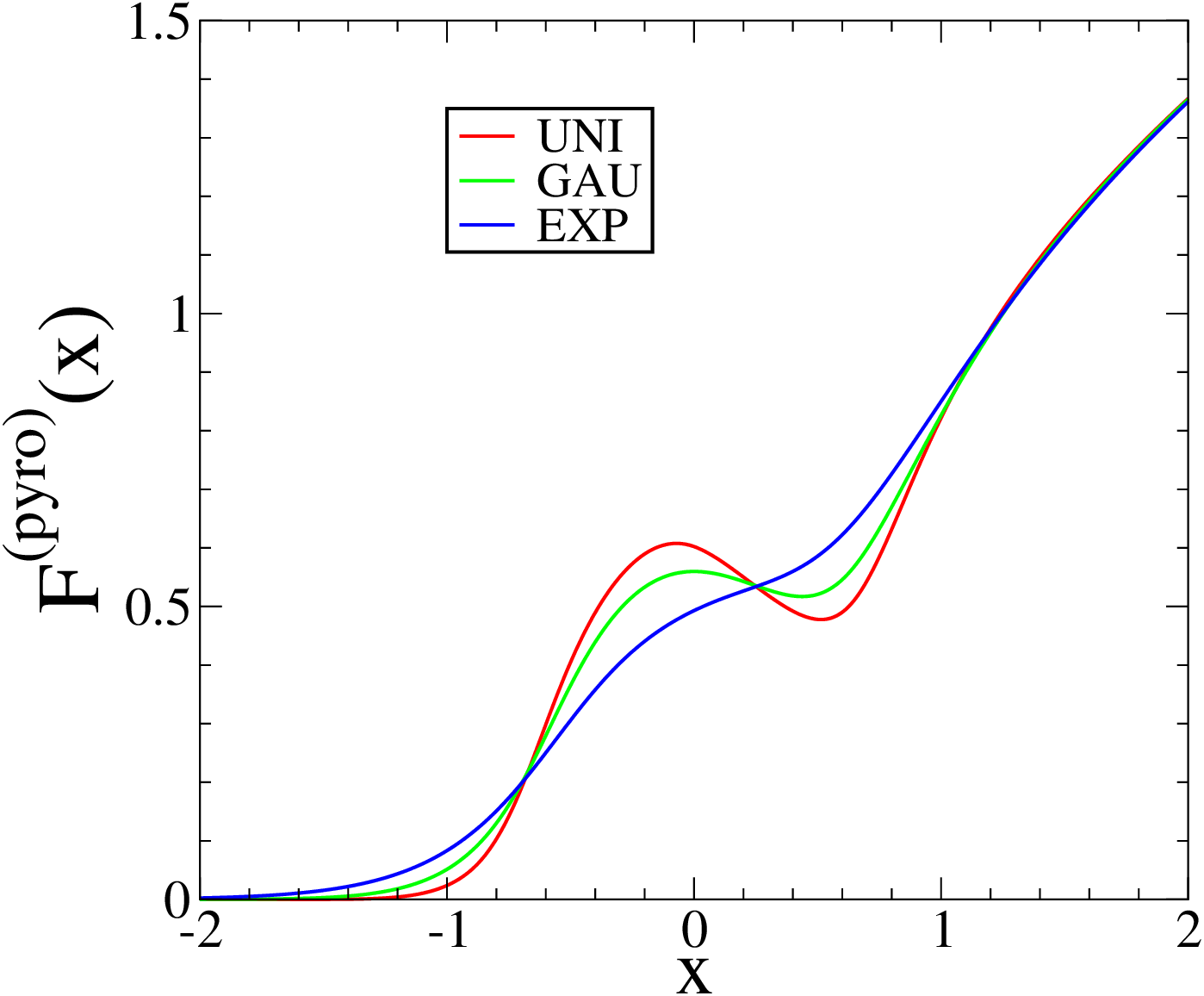}
\includegraphics[angle=0,width=.49\linewidth]{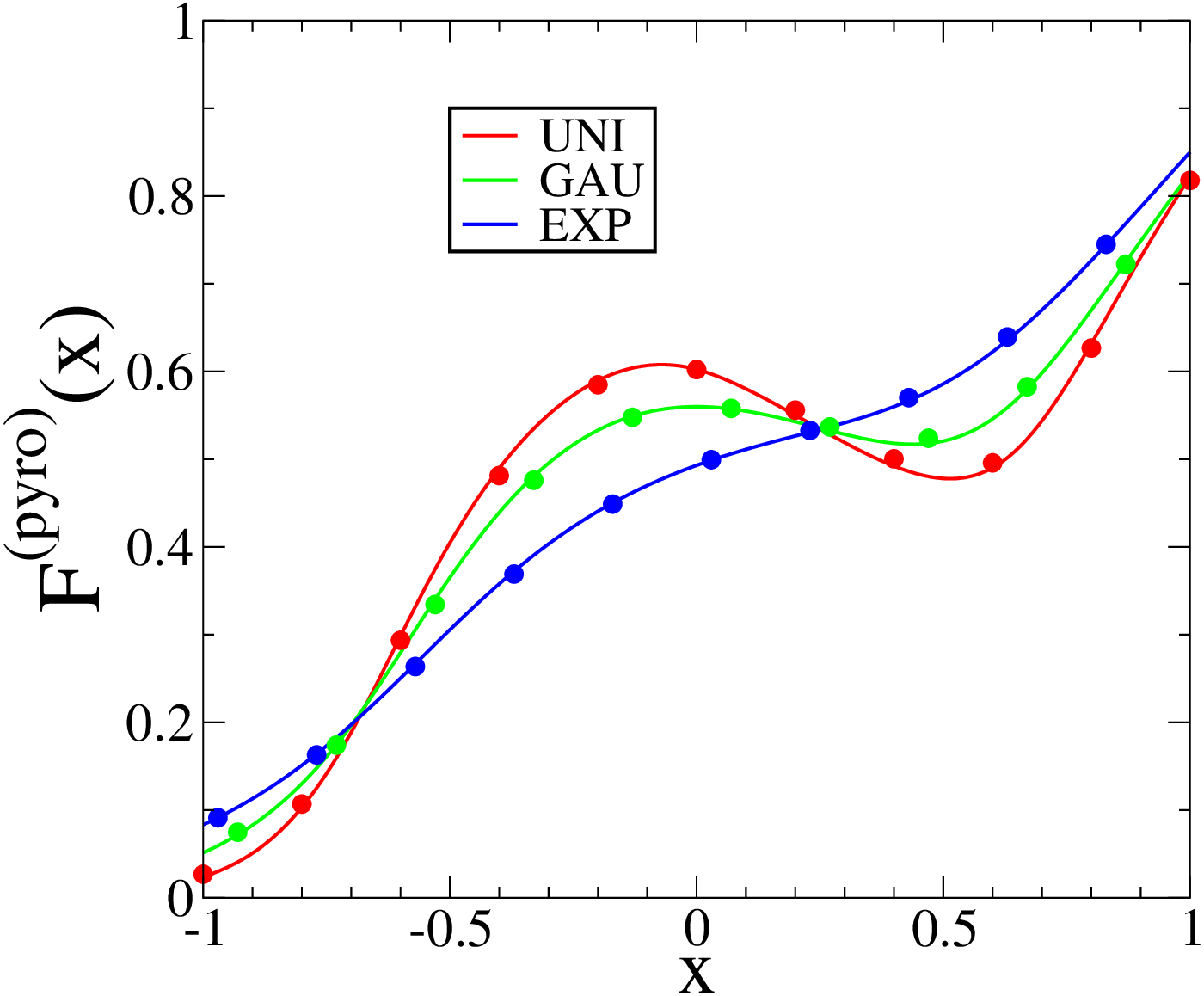}
\caption{\small
Left: the non-universal scaling function $F^\pyro(x)$
(see~(\ref{fpyro})) entering the prediction~(\ref{pyrores})
for the Lyapunov exponent of the pyrochlore ladder near $E=0$,
against reduced energy $x=E/w$, for three disorder distributions:
uniform, Gaussian and exponential.
Right: enlargement of the central part of the left plot.
Symbols: Numerical results for $w=0.001$.}
\label{gammapyro}
\end{center}
\end{figure}

The scaling function $F^\pyro(x)$ is plotted in figure~\ref{gammapyro}
for the uniform, Gaussian and exponential disorder distributions.
The right panel, showing an enlargement of the central part of the left one,
demonstrates that the analytical prediction~(\ref{pyrores})
is in excellent agreement with numerical results for the Lyapunov exponent (symbols),
based on products of $10^8$ matrices for each type disorder with $w=0.001$.

The most salient feature of these plots is that they are very asymmetric.
It is indeed expected that the Lyapunov exponent (inverse localization length)
is much smaller for $x$ negative, i.e., on the side of the dispersive band,
than for $x$ positive, i.e., on the side of the gap.
From a quantitative viewpoint, deep in the gap ($x\gg1$, i.e., $E\gg w$),
$g_2(x)$ falls off rapidly to zero, whereas $h_2(x)\approx x$ (see~(\ref{h2r})).
We thus obtain the universal law
\beq
F^\pyro(x)\approx\sqrt{x},
\label{fsqrt}
\eeq
irrespective of the disorder distribution.
This collapse is clearly seen in the right part of the left panel.
The formula~(\ref{fsqrt}) can be recast as $\gamma\approx\sqrt{E}$.
This estimate, which does not involve disorder at all,
is in agreement with the quadratic form of the lower dispersive band
near $E=0$, i.e., $E\approx-(\pi-q)^2$ (see~(\ref{sppyro})),
and so $\gamma=\abs{\Im q\,}\approx\sqrt{E}$.
For the uniform distribution,
$g_2(x)$ vanishes identically for $\abs{x}>\sqrt{3}$ (see~(\ref{unig2})).
As a consequence, $F^\pyro(x)$ vanishes for $x<-\sqrt{3}$,
implying that $\gamma$ vanishes more rapidly than $\sqrt{w}$
as $w$ tends to zero while $E<-w\sqrt{3}$.
The Lyapunov exponent is indeed expected to exhibit the usual scaling in $w^2$
at any fixed energy inside the dispersive band.
In the central part, enlarged in the right panel, the scaling function $F^\pyro(x)$
exhibits a rather strong dependence on the disorder distribution.
It is non-monotonic for the uniform and Gaussian distributions.

\section{The diamond chain}
\label{secdiam}

\begin{figure}[!ht]
\begin{center}
\includegraphics[angle=0,width=.45\linewidth]{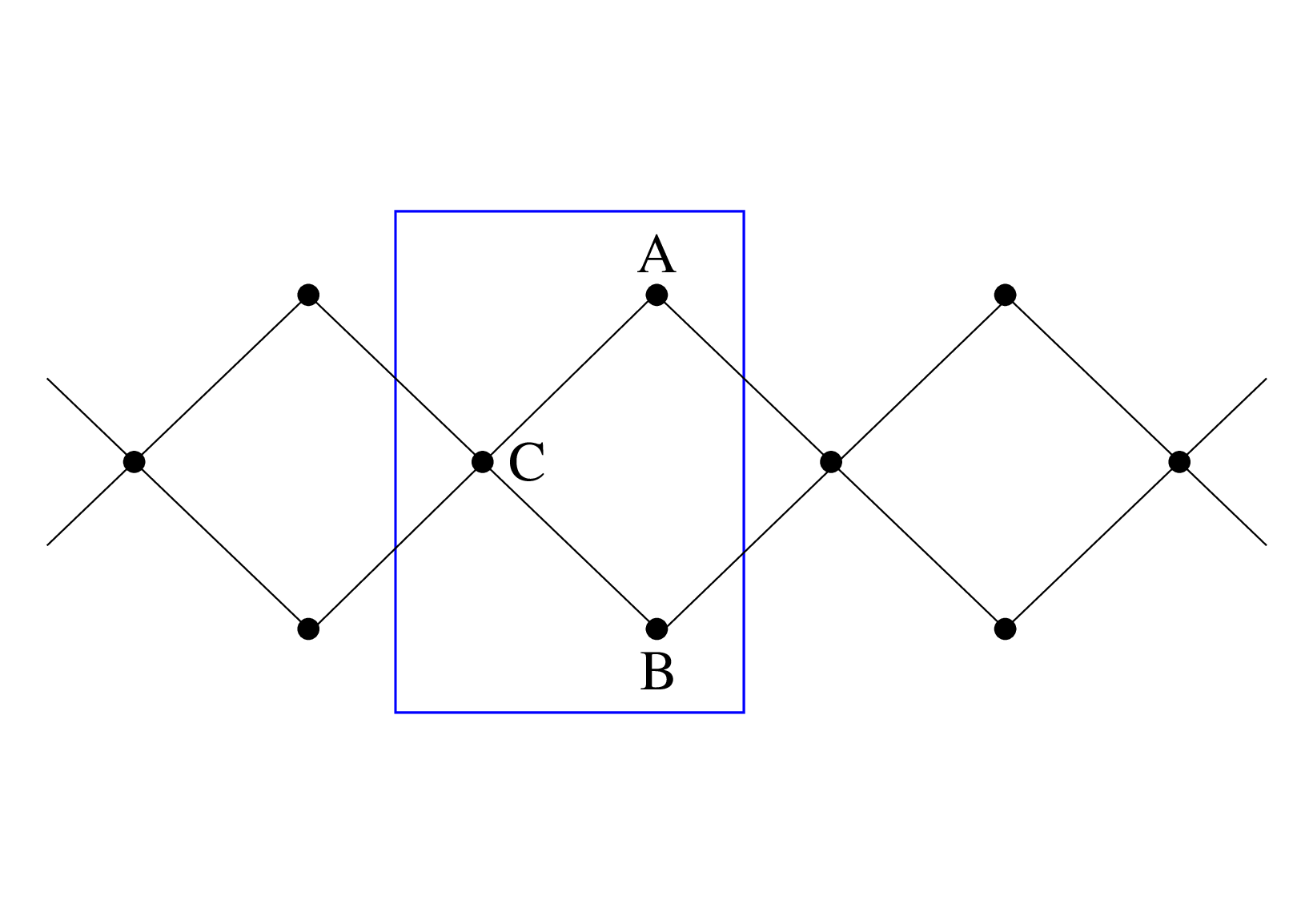}
\hskip 10pt
\includegraphics[angle=0,width=.35\linewidth]{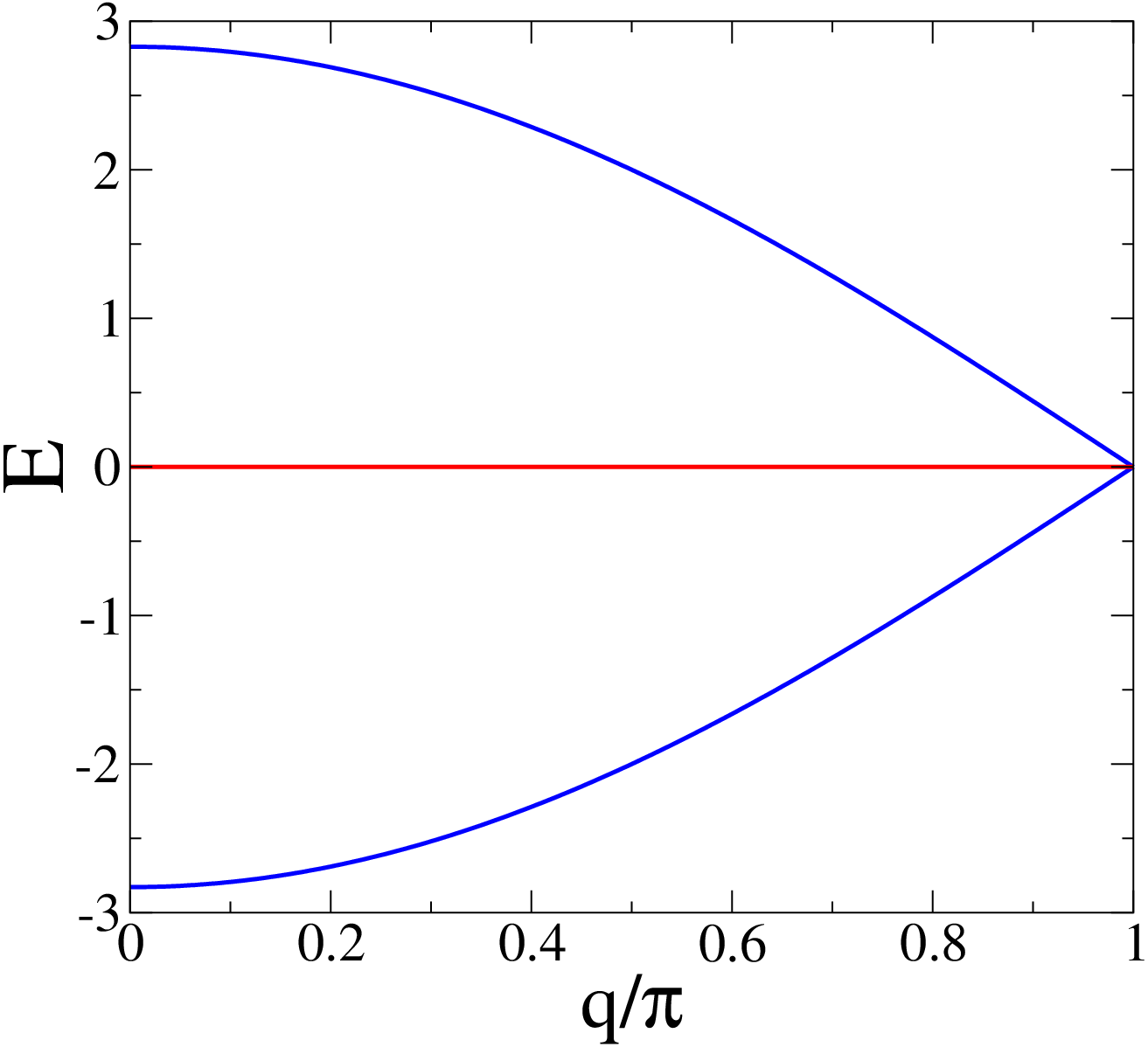}
\caption{\small
Left: the diamond chain has 3 atoms per unit cell.
Right: its spectrum (see~(\ref{spdiam})) has one flat band (red)
and two dispersive ones (blue).}
\label{diam}
\end{center}
\end{figure}

The diamond chain (see figure~\ref{diam}, left)
is another 1D structure known to exhibit a flat band~\cite{FL1,FL3}.
We equip it with a tight-binding model with diagonal disorder.
All disordered potentials are again drawn from the same distribution.
The tight-binding equations at energy $E$ read
\beqa
(E-v_n^A)\psi_n^A&=&\psi_n^C+\psi_{n+1}^C,
\nonumber\\
(E-v_n^B)\psi_n^B&=&\psi_n^C+\psi_{n+1}^C,
\nonumber\\
(E-v_n^C)\psi_n^C&=&\psi_{n-1}^A+\psi_{n-1}^B+\psi_n^A+\psi_n^B.
\label{tbdiam}
\eeqa

The dispersion relation of a clean chain
is obtained by expressing that $E$ is an eigenvalue of the matrix
\beq
M(q)=\pmatrix{0&0&1+\e^{\ii q}\cr 0&0&1+\e^{\ii q}\cr
1+\e^{-\ii q}&1+\e^{-\ii q}&0}.
\eeq
The spectrum (see figure~\ref{diam}, right)
consists of one flat band and two dispersive ones:
\beqa
E&=&0,
\nonumber\\
E&=&\pm2\sqrt{2}\cos\frac{q}{2}.
\label{spdiam}
\eeqa
Both dispersive bands intersect the flat one at the edges of the Brillouin zone ($q=\pm\pi)$.
The flat band consist of the extensively degenerate states
\beqa
\psi_n^A=-\psi_n^B=\sigma_n,
\label{degenediam}
\eeqa
where $\sigma_n$ are arbitrary amplitudes.
The most localized states, corresponding to e.g.~$\sigma_n=\delta_{n0}$, live on two sites.

The weakly disordered flat band of the diamond chain
has been investigated recently~\cite{FL1,FL3}.
Here, too, as dispersive bands intersect the flat band,
the localization length is expected to be large when the disorder is weak.
From a quantitative viewpoint,
a scaling of the localization length in $w^{-4/3}$ was predicted in References~\cite{FL1,FL3}.
The subsequent analysis refines this finding.
We shall indeed demonstrate that the Lyapunov exponent obeys two different scaling laws
in two successive regimes as energy goes away from the pristine flat band,
scaling first as~$w^{4/3}$ in Regime~I ($E\sim w^{5/3}$)
and then as $w$ in Regime~II ($E\sim w$).

The scaling analysis of the model goes as follows.
Introducing the combinations
\beqa
s_n&=&\psi_n^A+\psi_n^B,
\nonumber\\
t_n&=&\psi_n^A-\psi_n^B,
\eeqa
one can eliminate $t_n$ from the tight-binding equations~(\ref{tbdiam}),
whereas the remaining equations can be recast -- without any approximation -- as
\beq
\pmatrix{s_n\cr\psi_{n+1}^C}=T_n\pmatrix{s_{n-1}\cr\psi_n^C},
\eeq
where
\beq
T_n=-\pmatrix{1&-\eta_n^C\cr\chi_n & 1-\chi_n\eta_n^C},
\label{tdiam}
\eeq
with
\beq
\eta_n^A=E-v_n^A,\quad
\eta_n^B=E-v_n^B,\quad
\eta_n^C=E-v_n^C
\eeq
and
\beq
\chi_n=\frac{\eta_n^A\eta_n^B}{\eta_n^A+\eta_n^B}.
\eeq
The above non-linear combination of disorder variables
again has a broad probability distribution with symmetric Cauchy tails (see~\ref{appcau}).
It is therefore charac\-terized by its Cauchy strength and its mean value:
\beqa
G_1(E)&=&\bigmean{\delta\left(\frac{1}{\chi_n}\right)}=
\bigmean{\delta\left(\frac{\eta_n^A+\eta_n^B}{\eta_n^A\eta_n^B}\right)},
\nonumber\\
H_1(E)&=&\bigmean{\chi_n}=\bigmean{\frac{\eta_n^A\eta_n^B}{\eta_n^A+\eta_n^B}}.
\label{g1h1}
\eeqa
These quantities are investigated in~\ref{appdiam}.
Their scaling dependence on $w$ reads (see~(\ref{g1h1app}))
\beq
G_1(E)=wg_1(x),\qquad H_1(E)=wh_1(x),\qquad x=\frac{E}{w}.
\label{g1h1sca}
\eeq
The functions $G_1(E)$ and $g_1(x)$ are even, whereas $H_1(E)$ and $h_1(x)$ are odd.

The transfer matrix~(\ref{tdiam}) obeys $\det T_n=1$.
The associated Riccati variables obey the recursion
\beq
R_n=\chi_n+\frac{R_{n-1}}{1-\eta_n^CR_{n-1}},
\label{rdiam}
\eeq
where the Cauchy distributed random quantity $\chi_n$ enters additively.
Equivalently, their reciprocals $S_n=1/R_n$ obey
\beq
S_n=\frac{S_{n-1}-\eta_n^C}{1+\chi_n(S_{n-1}-\eta_n^C)}.
\label{ydiam}
\eeq
The associated Lyapunov exponent reads
\beq
\gamma=\mean{\ln\abs{1-\eta_n^C R_{n-1}}}.
\label{gdiam}
\eeq
It gives the inverse localization length, in units of the size of a unit cell.

It is worth beginning with a heuristic scaling analysis.
Consider first the situation where $E=0$ exactly.
One can get an idea of the dynamics generated by the recursion~(\ref{rdiam})
at weak disorder by comparing two extreme initial conditions.
If the Riccati variables are initialized to $R_0=0$,
(\ref{rdiam}) yields $R_n\approx R_{n-1}+\chi_n$,
so that $R_n$ is the sum of $n$ Cauchy variables, hence $R_n\sim nw$.
Setting $R_0=\infty$, i.e., $S_0=0$,
(\ref{ydiam}) yields $S_n\approx S_{n-1}+\eta_n^C$, with $\mean{\eta_n^C}=E=0$,
and so $S_n$ is the sum of~$n$ random variables with zero mean and finite variance,
hence $S_n\sim n^{1/2}w$.
Equating both estimates yields $R_\star\sim w^{-1/3}$.
This is the predicted scaling of the typical value of the Riccati variable.
Furthermore, the number of steps needed to relax to this typical scale
from either initial condition scales as $n_\star\sim w^{-4/3}$.
This provides an estimate of the correlation length.
We thus obtain $\gamma\sim w^{4/3}$.
At non-zero energy $E$,
the above scaling laws hold as long as the cumulative effect of $\mean{\eta_n^C}$
does not take the lead.
This condition reads $En_\star R_\star\sim1$, i.e., $E\sim w^{5/3}$, or $x\sim w^{2/3}$.
We shall refer to this range as Regime~I.
Consider now Regime~II, defined as $E\sim w$, or $x\sim1$.
The dynamics with $R_0=0$ is unchanged,
whereas the initial condition $S_0=0$ yields $S_n\sim nE$.
Equating both estimates yields $R_\star\sim1$ and $\gamma\sim w$.

The quantitative analysis of the weak-disorder regime exposed below
corroborates the existence of two successive scaling regimes.
The random recursion~(\ref{rdiam}) can be expanded as
\beq
R_n-R_{n-1}\approx\chi_n+\eta_n^C R_{n-1}^2+(\eta_n^C)^2R_{n-1}^3,
\label{recr}
\eeq
where each term is small, and higher-order terms are negligible.
The distribution of~$\chi_n$ has been described above (see~(\ref{g1h1}),~(\ref{g1h1sca})).
The distribution of $\eta_n^C$ is a narrow regular one,
with $\mean{\eta_n^C}=E=wx$
and $\mean{(\eta_n^C)^2}=E^2+w^2=w^2(1+x^2)$.
As a consequence, the invariant distribution $f(R)$ of the Riccati variable obeys
\beqa
f(R)\approx
&-&wh_1(x)\frac{\dd}{\dd R}f(R)
+wg_1(x)\int_{-\infty}^\infty\frac{f(R-\chi)\,\dd\chi}{\chi^2+(\pi wg_1(x))^2}
\nonumber\\
&-&wx\frac{\dd}{\dd R}(R^2f(R))+\frac{w^2}{2}\frac{\dd^2}{\dd R^2}(R^4f(R))
\nonumber\\
&-&w^2(1+x^2)\frac{\dd}{\dd R}(R^3f(R)).
\label{freq}
\eeqa
The three lines correspond to the three contributions in the r.h.s.~of~(\ref{recr}), in the same order.
In the third line, the term involving a second derivative has been skipped as it is negligible.

In order to solve the integro-differential equation~(\ref{freq}),
it is advantageous to introduce the Hilbert transform
\beq
L(p)=\int_{-\infty}^\infty\frac{f(R)\,\dd R}{p+\ii R},
\label{ldef}
\eeq
where $p$ is a complex variable obeying $\Re p>0$.
The use of Hilbert transforms in the analysis of one-dimensional disordered systems
dates back to Dyson's pioneering work~\cite{Dy}.
It has been instrumental in a range of other
investigations~\cite{cltt,tn1,nl,bl,li}.
In the present situation,
the Hilbert transform of the second term in the first line on the r.h.s.~of~(\ref{freq})
is simply $L(p+\pi wg_1(x))\approx L(p)+\pi wg_1(x)L'(p)$,
where accents denote derivatives w.r.t.~$p$.
Integrations by parts readily yield the Hilbert transforms of all other terms.
We are left with
\beqa
0&\approx&w(\ii h_1(x)+\pi g_1(x))L'(p)
-\ii wx(p^2L'(p)+2pL(p)-1)
\nonumber\\
&+&\frac{w^2}{2}(-p^4L''(p)-8p^3L'(p)-12p^2L(p)+6p)
\nonumber\\
&+&w^2(1+x^2)(p^3L'(p)+3p^2L(p)-2p),
\label{leq}
\eeqa
where terms are written in the same order as in~(\ref{freq}).
The above equation will be solved successively in both regimes
evidenced by the heuristic analysis.

\subsubsection*{Regime~I ($E\sim w^{5/3}$).}

This first regime can be parametrized by the scaling variable
\beq
y=\frac{E}{w^{5/3}}=\frac{x}{w^{2/3}}.
\eeq
The reduced variable $x$ is small, and so $g_1(x)$ can be approximated by $g_1(0)$,
while $h_1(0)$ vanishes by symmetry.
Setting
\beq
\lambda=\frac{w^{1/3}}{(2\pi g_1(0))^{1/3}},\qquad
Y=\frac{2^{2/3}}{(\pi g_1(0))^{1/3}}\,y,
\label{lYdef}
\eeq
and
\beq
p=\frac{s}{\lambda},\qquad
L(p)=\lambda M(s),
\eeq
all terms in the r.h.s.~of~(\ref{leq}) contribute on the same footing,
confirming the relevance of the scaling $E\sim w^{5/3}$.
We are left with a dimensionless second-order differential equation for $M(s)$,
which can be integrated once, yielding
\beq
s^4M'(s)+(2s^3+\ii Ys^2-1)M(s)=s^2+\ii Ys-\phi(Y).
\label{meq}
\eeq

The complex function $\phi(Y)$,
which entered~(\ref{meq}) as an integration constant, is of central importance.
It will indeed determine the universal scaling function $F^\I(Y)$
involved in the result~(\ref{gI}) for the Lyapunov exponent.
It can be determined as follows (see~\cite{li} and~\cite{cltt}).
Using the complex variable $u=1/s$ instead of $s$,
and setting $\wt M(u)=M(s)$, the general solution to~(\ref{meq}) reads
\beq
\wt M(u)=u^2\e^{\ii Yu-u^3/3}K(u),
\eeq
with
\beq
K'(u)=-\left(\frac{1}{u^2}+\frac{\ii Y}{u}-\phi(Y)\right)\e^{-\ii Yu+u^3/3}.
\label{kprime}
\eeq
Consider the behavior of $\wt M(u)$ for $u=\rho\,\e^{\pm\ii\pi/3}$,
with $\rho$ real and positive, so that $u^3=-\rho^3$.
In the $\rho\to+\infty$ limit in either direction,
the growth of the exponential factor in the expression of $\wt M(u)$
must be compensated by an exponential falloff of $K(u)$.
We thus obtain the condition
$K(\e^{\ii\pi/3}\infty)=K(\e^{-\ii\pi/3}\infty)=0$, hence
\beq
\int_\C K'(u)\dd u=0,
\eeq
where the contour $\C$ joins $\e^{-\ii\pi/3}\infty$ to $\e^{\ii\pi/3}\infty$.
An integration by parts yields
\beq
\int_\C(\phi(Y)-u)\e^{-\ii Yu+u^3/3}\dd u=0.
\eeq
Recognizing the integral representations of the Airy function and of its derivative, we obtain
\beq
\phi(Y)=-\frac{\Ai'(\ii Y)}{\Ai(\ii Y)}.
\label{phires}
\eeq

The scaling behavior of the Lyapunov exponent
can be investigated by averaging the expression~(\ref{gdiam})
first over the invariant distribution of the Riccati variable $R_{n-1}$,
then over the disorder $\eta_n^C$.
The first average goes as follows.
The definition~(\ref{ldef}) implies
\beq
\Lambda(p)
=\bigmean{\ln\left(1+\frac{\ii R}{p}\right)}
=\int_p^\infty\left(\frac{1}{q}-L(q)\right)\dd q.
\eeq
Only the first two terms of the large-$p$ expansion of $\Lambda(p)$ will be needed.
The differential equation~(\ref{kprime}) yields the expansion
\beq
K(u)=\frac{1}{u}+K_0(Y)+\left(\phi(Y)-\frac{Y^2}{2}\right)u+\cdots\qquad(u\to0),
\eeq
where the explicit value of the complex integration constant $K_0(Y)$ will not be needed.
The above expansion translates into
\beq
\Lambda(p)=-\frac{\ii Y+K_0(Y)}{\lambda p}+\frac{Y^2-\ii Y K_0(Y)-\phi(Y)}{2\lambda^2p^2}+\cdots
\eeq
Performing the substitution $p\to\ii/\eta_n^C$ in the above expansion,
taking the real part and averaging over $\eta_n^C$,
we obtain the following prediction for the scaling behavior of the Lyapunov exponent all over Regime~I:
\beq
\gamma\approx\frac{(\pi g_1(0))^{2/3}}{2^{1/3}}\,F^\I(Y)\;w^{4/3},
\label{gI}
\eeq
where
\beq
F^\I(Y)=\Re\phi(Y),
\label{FI}
\eeq
with $Y$ defined in~(\ref{lYdef}) and $\phi(Y)$ given by~(\ref{phires}).

The result~(\ref{gI}) depends on energy through the universal scaling function $F^\I(Y)$.
It depends on the disorder distribution through $g_1(0)$,
which enters both~(\ref{gI}) explicitly
and the definition~(\ref{lYdef}) of the scaling variable $Y$.
As predicted by the heuristic reasoning, the Lyapunov exponent scales as $w^{4/3}$,
whereas the typical Riccati variable scales as $1/\lambda$, i.e., as $w^{-1/3}$.

The scaling function $F^\I(Y)$ is even.
Its behavior at small $Y$ reads
\beq
F^\I(Y)=f_0+f_2Y^2+\cdots,
\eeq
with
\beqa
f_0&=&-\frac{\Ai'(0)}{\Ai(0)}=\frac{3^{1/3}\Gamma(2/3)}{\Gamma(1/3)}=0.729011,
\nonumber\\
f_2&=&\frac{1}{2}+\left(\frac{\Ai'(0)}{\Ai(0)}\right)^3
=\frac{1}{2}-3\left(\frac{\Gamma(2/3)}{\Gamma(1/3)}\right)^3=0.112561.
\eeqa
Its behavior at large $Y$,
\beq
F^\I(Y)\approx\left(\frac{\abs{Y}}{2}\right)^{1/2},
\eeq
yields the estimate
\beq
\gamma\approx\left(\frac{\pi g_1(0)\abs{E}w}{2}\right)^{1/2},
\label{largeI}
\eeq
in agreement with the small-$x$ behavior in Regime~II (see~(\ref{smallII})).

\begin{figure}[!ht]
\begin{center}
\includegraphics[angle=0,width=.49\linewidth]{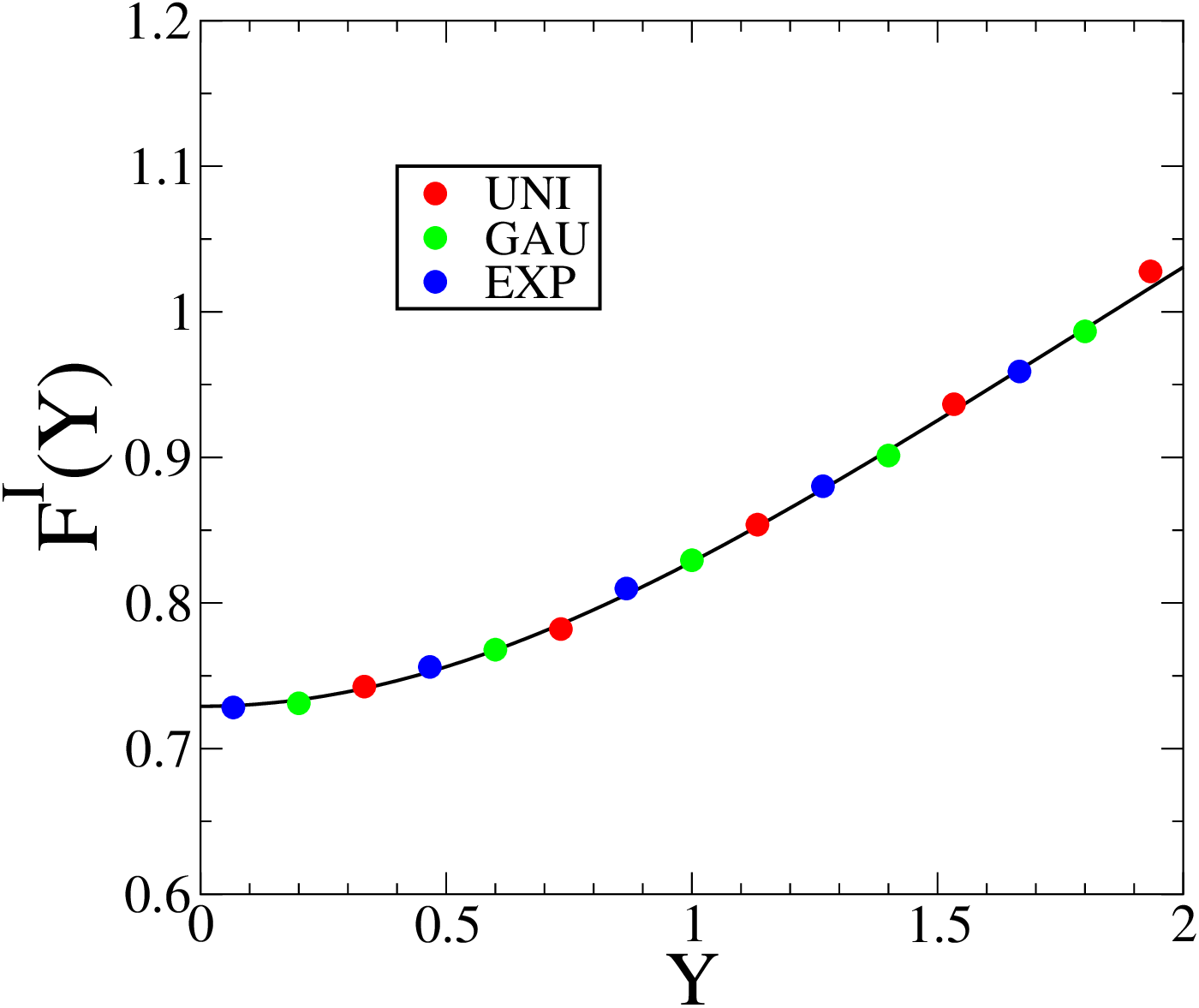}
\caption{\small
The universal scaling function $F^\I(Y)$ entering the prediction~(\ref{gI})
for the Lyapunov exponent of the diamond chain in Regime~I,
against the rescaled energy variable $Y$.
Symbols: Numerical results for three disorder distributions:
uniform, Gaussian and exponential, with $w=0.0025$.}
\label{gamma1diam}
\end{center}
\end{figure}

The scaling function $F^\I(Y)$ entering the prediction~(\ref{gI})
is plotted in figure~\ref{gamma1diam}.
A good agreement is observed with numerical results for the Lyapunov exponent (symbols),
based on products of $10^9$ matrices
for uniform, Gaussian and exponential disorders with $w=0.0025$.
The corresponding values of $g_1(0)$ are given in table~\ref{tabdis}.

\subsubsection*{Regime~II ($E\sim w$).}

This second regime is parametrized by the scaling variable
\beq
x=\frac{E}{w}.
\eeq
The heuristic reasoning predicts that typical Riccati variables are of order unity.
As a consequence, no rescaling of the variable $p$ is involved in the analysis of~(\ref{leq}).
Keeping only terms proportional to $w$, the latter equation boils down to
\beq
(p^2-\mu(x)^2)L'(p)+2pL(p)=1,
\label{leq2}
\eeq
where we have introduced the complex function
\beq
\mu(x)=\alpha(x)-\ii\beta(x)=\left(\frac{h_1(x)-\ii\pi g_1(x)}{x}\right)^{1/2},
\eeq
hence
\beqa
\alpha(x)=\frac{1}{\abs{x}\sqrt{2}}\left(\abs{x}\sqrt{h_1(x)^2+\pi^2g_1(x)^2}+xh_1(x)\right)^{1/2},
\nonumber\\
\beta(x)=\frac{1}{x\sqrt{2}}\left(\abs{x}\sqrt{h_1(x)^2+\pi^2g_1(x)^2}-xh_1(x)\right)^{1/2}.
\eeqa
We recall that $g_1(x)$ is even and $h_1(x)$ is odd,
and so $\alpha(x)$ is even and $\beta(x)$ is odd.
The solution to~(\ref{leq2}) is
\beq
L(p)=\frac{1}{p+\mu(x)}.
\eeq
There can indeed be no pole at $p=\mu(x)$, whose real part $\alpha(x)$ is positive.
The corresponding invariant distribution of the Riccati variable is Cauchy:
\beq
f(R)=\frac{\alpha(x)}{\pi\left((R+\beta(x))^2+\alpha(x)^2\right)}.
\eeq

The scaling behavior of the Lyapunov exponent is obtained by linearizing
the expression~(\ref{gdiam}).
We thus obtain $\gamma\approx-\mean{\eta^C}\mean{R}$,
with $\mean{\eta^C}=E=xw$ and $\mean{R}=-\beta(x)$, by symmetry, and so
\beq
\gamma\approx w\,F^\II(x),
\label{gII}
\eeq
in qualitative agreement with the heuristic reasoning.
The corresponding scaling function,
\beq
F^\II(x)=x\beta(x)
=\frac{1}{\sqrt{2}}\left(\abs{x}\sqrt{h_1(x)^2+\pi^2g_1(x)^2}-xh_1(x)\right)^{1/2},
\label{FII}
\eeq
is non-universal, as it depends on the full shape of the disorder distribution,
through the functions $g_1(x)$ and $h_1(x)$, investigated in~\ref{apppyro}.

The function $F^\II(x)$ is even, as should be.
For small values of $x$, we have
\beq
F^\II(x)\approx\left(\frac{\pi g_1(0)\abs{x}}{2}\right)^{1/2},
\label{smallII}
\eeq
so that a smooth crossover to Regime~I is observed (see~(\ref{largeI})).

\begin{figure}[!ht]
\begin{center}
\includegraphics[angle=0,width=.49\linewidth]{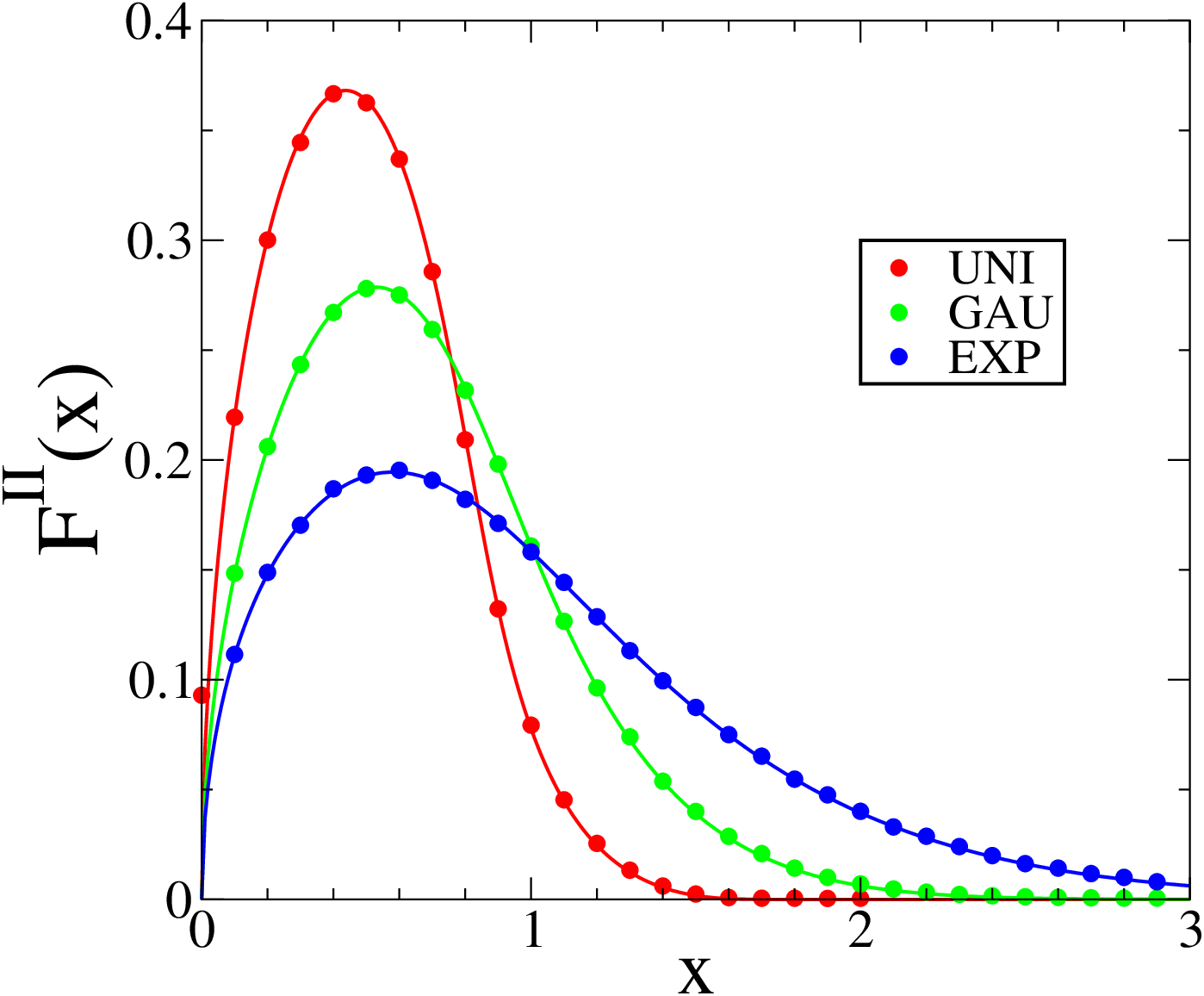}
\caption{\small
The non-universal scaling functions $F^\II(x)$ entering the pre\-diction~(\ref{gII})
for the Lyapunov exponent of the diamond chain in Regime~II,
against the rescaled energy variable $x$,
for three disorder distributions: uniform, Gaussian and exponential.
Symbols: numerical results for $w=0.0025$.}
\label{gamma2diam}
\end{center}
\end{figure}

For the uniform, Gaussian and exponential distributions,
explicit expressions of~$g_1(x)$ are given in~\ref{appdis},
whereas $h_1(x)$ is expressed as an integral
that can be carried out in closed form only for Gaussian disorder.
The resulting scaling functions $F^\II(x)$ are plotted in figure~\ref{gamma2diam}.
The prediction~(\ref{gII}) is found in excellent agreement with
numerical results for the Lyapunov exponent (symbols),
based on products of $10^8$ matrices for each type disorder, with $w=0.0025$.
For the uniform distribution,
$g_1(x)$ vanishes identically for $\abs{x}>\sqrt{3}$ (see~(\ref{unig1})).
As a consequence, $F^\II(x)$ also vanishes there,
implying that $\gamma$ falls off more rapidly than linearly in $w$ in that range.

Gathering our findings on both scaling regimes at weak disorder,
we predict that the Lyapunov exponent is a highly non-monotonic function of energy.
Considering only positive energies for definiteness,
$\gamma$ exhibits a minimum at the center of the dispersive band ($E=0$),
scaling as $w^{4/3}$ (Regime~I),
a maximum for $E\sim w$, scaling as $w$ (Regime~II),
a second minimum near the middle of the dispersive band, scaling as $w^2$,
before it reaches values of order unity if $E$ exceeds the outer band edge ($E>2\sqrt2$)
and then keeps growing indefinitely.
The highly non-monotonic profile sketched above
is a weak-disorder feature, which only takes place below some non-universal threshold.
The minimum at the band center only occurs below
$w_c\approx0.32$ for the uniform distribution,
and below $w_c\approx0.90$ for the Gaussian distribution,
with the conventions used in this work.
This explains why it is not visible on the data shown in Reference~\cite{FL1}.
For the exponential distribution, the maximum for $E\sim w$ disappears at $w_c\approx1.69$,
and so the Lyapunov exponent increases monotonically with energy whenever $w$ is above $w_c$.

\section{Discussion}
\label{disc}

We have investigated the scaling behavior of the localization length
in the weak-disorder regime of various 1D structures having flat bands.
The foremost feature is the power-law divergence
$\xi\sim w^{-\nu}$ of the localization length at weak disorder.
Earlier studies~\cite{FL1,FL2,FL3,FL4,Ge} have shown that
the two well-known values of the exponent~$\nu$ pertaining to the Anderson problem on the chain,
\beq
\nu=2/3,\quad 2,
\eeq
are supplemented by a few more more exotic values
\beq
\nu=0,\quad 1/2,\quad 1,\quad 4/3,
\label{exos}
\eeq
pertaining to 1D flat-band systems.
To our knowledge, whether the above list is complete or whether it can be derived
by systematic reasoning are still open questions.

The main focus of the present work has been on the full scaling functions
describing the dependence of the Lyapunov exponent (inverse localization length)~$\gamma$
on energy $E$ and disorder width $w$ throughout weakly disordered flat bands.
A careful distinction has been made between situations where the scaling functions are universal
(i.e., depend on the disorder distribution only through its width)
and where they depend on the shape of the disorder distribution.
Three examples
(stub chain, pyrochlore ladder, diamond chain),
exhausting the full list~(\ref{exos}) of exotic exponents known to date,
have been analyzed in detail.
In each case, the analysis of the localization length at weak disorder
has been reduced to investigating the Lyapunov exponent of products
of random $2\times2$ transfer matrices.
The latter reduction,
which is asymptotically exact to leading order in the weak-disorder regime,
can be thought of as a thorough implementation of the detangling procedure
put forward in Reference~\cite{FL2}.

Our main specific findings are as follows.
On the stub chain (section~\ref{secstub}),
where one central flat band is isolated from two lateral dispersive ones,
the localization length remains microscopic at weak disorder.
Its behavior is only accessible by perturbative and numerical means.
Its non-universal dependence on the ratio $x=E/w$ exhibits several features of interest
(see figure~\ref{gammastub} and table~\ref{tabstub}).
In particular, the Lyapunov exponent takes a constant value $\gamma_\infty$
(see~(\ref{ginf})) for $\abs{x}\gg1$, i.e., $w\ll\abs{E}\ll1$.
The latter constant is dictated by the band structure of the pristine system.
It is formally infinite in situations where all bands are flat.
An interesting example is provided by the diamond ladder subjected to a magnetic field,
in the case where each unit cell is pierced by half a flux quantum~\cite{VDMB,DV}.
The spectrum of that system consists of three flat bands at energies $E=0$, $\pm2$.
There, omitting details, we predict a logarithmic divergence of the Lyapunov exponent
of the form $\gamma\approx\ln\abs{x}+K$ for $\abs{x}\gg1$,
where~$K$ is an explicit non-universal constant,
depending on the full shape of the disorder distribution.

In the other geometries considered here,
the flat bands are not isolated from the dispersive ones,
and so the localization length diverges in the weak-disorder regime.
On the pyrochlore ladder (section~\ref{secpyro}),
the two flat bands are tangent to a dispersive one.
The localization length has scaling exponent 1/2
and obeys the non-universal scaling law~(\ref{pyrores}).
The dependence of the scaling function $F^\pyro(x)$
on the disorder distribution is given by~(\ref{fpyro})
in terms of the functions $g_2(x)$ and $h_2(x)$,
investigated in~\ref{apppyro} and
characteristic of the Cauchy effective disorder.
On the diamond chain (section~\ref{secdiam}),
a central flat band intersects two symmetric dispersive ones.
The situation at weak disorder is richer than the previous cases.
The localization length now exhibits two successive scaling regimes.
In Regime~I ($E\sim w^{5/3}$),
it diverges with exponent~4/3, according to~(\ref{gI}),
and a universal scaling function $F^\I(Y)$ given by~(\ref{FI}).
In Regime~II ($E\sim w$),
it diverges with exponent 1, according to~(\ref{gII}),
and a non-universal scaling function $F^\II(x)$ (see~(\ref{FII}))
depending on the full shape of the disorder distribution,
through the functions $g_1(x)$ and $h_1(x)$ investigated in~\ref{appdiam}.
Here, too, the lack of universality in this second regime
originates in the occurrence of a broad probability distribution
of the effective disorder, exhibiting symmetric Cauchy tails.

\ack

It is a pleasure to thank Alexei Andreanov,
Sergej Flach and Julien Vidal for useful correspondence.

\appendix

\section{Cauchy tails}
\label{appcau}

Cauchy tails in probability distributions
play a central part in the analysis
of the pyrochlore ladder (section~\ref{secpyro})
and the diamond chain (section~\ref{secdiam}).

\begin{figure}[!ht]
\begin{center}
\includegraphics[angle=0,width=.4\linewidth]{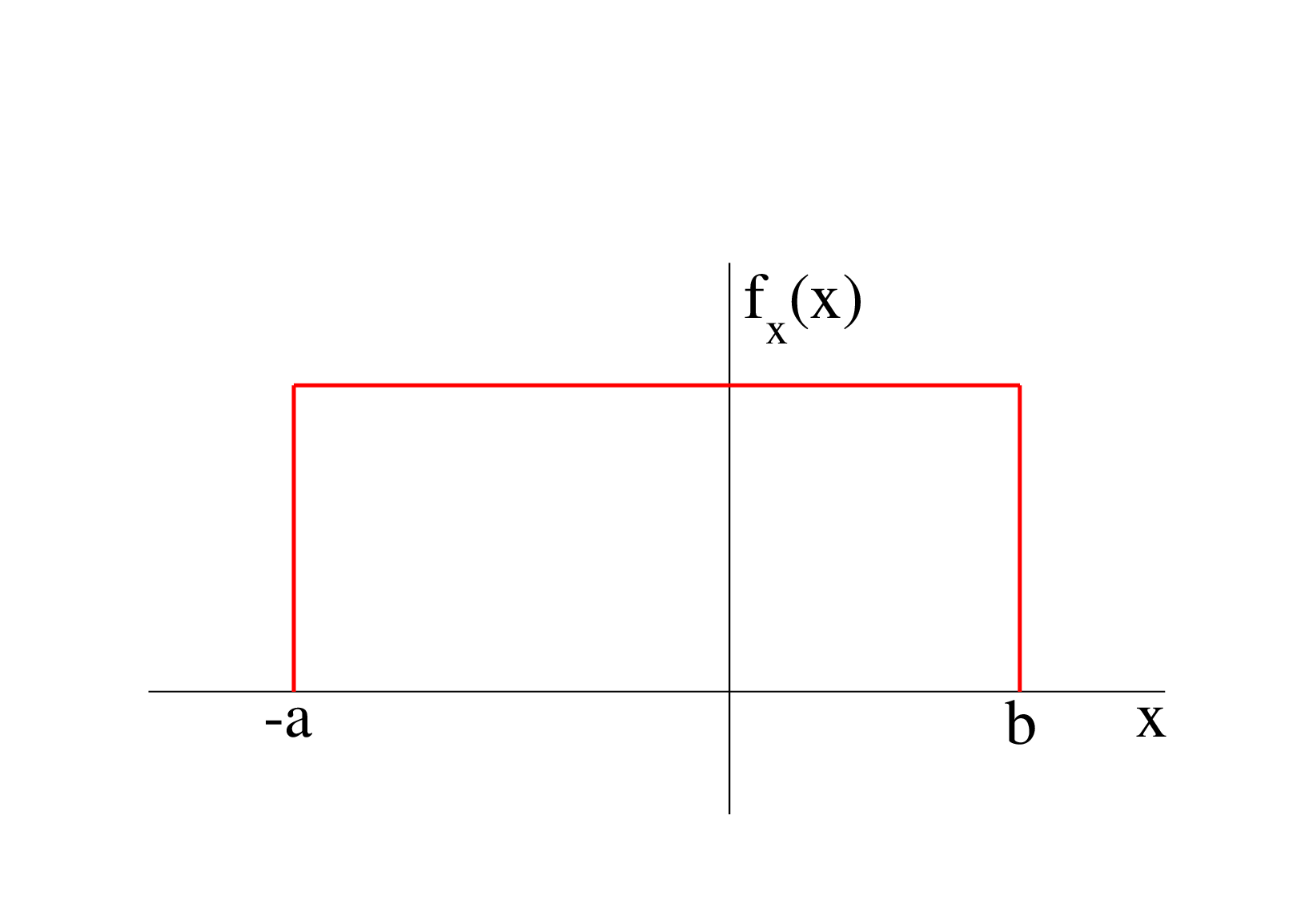}
\hskip 5pt
\includegraphics[angle=0,width=.4\linewidth]{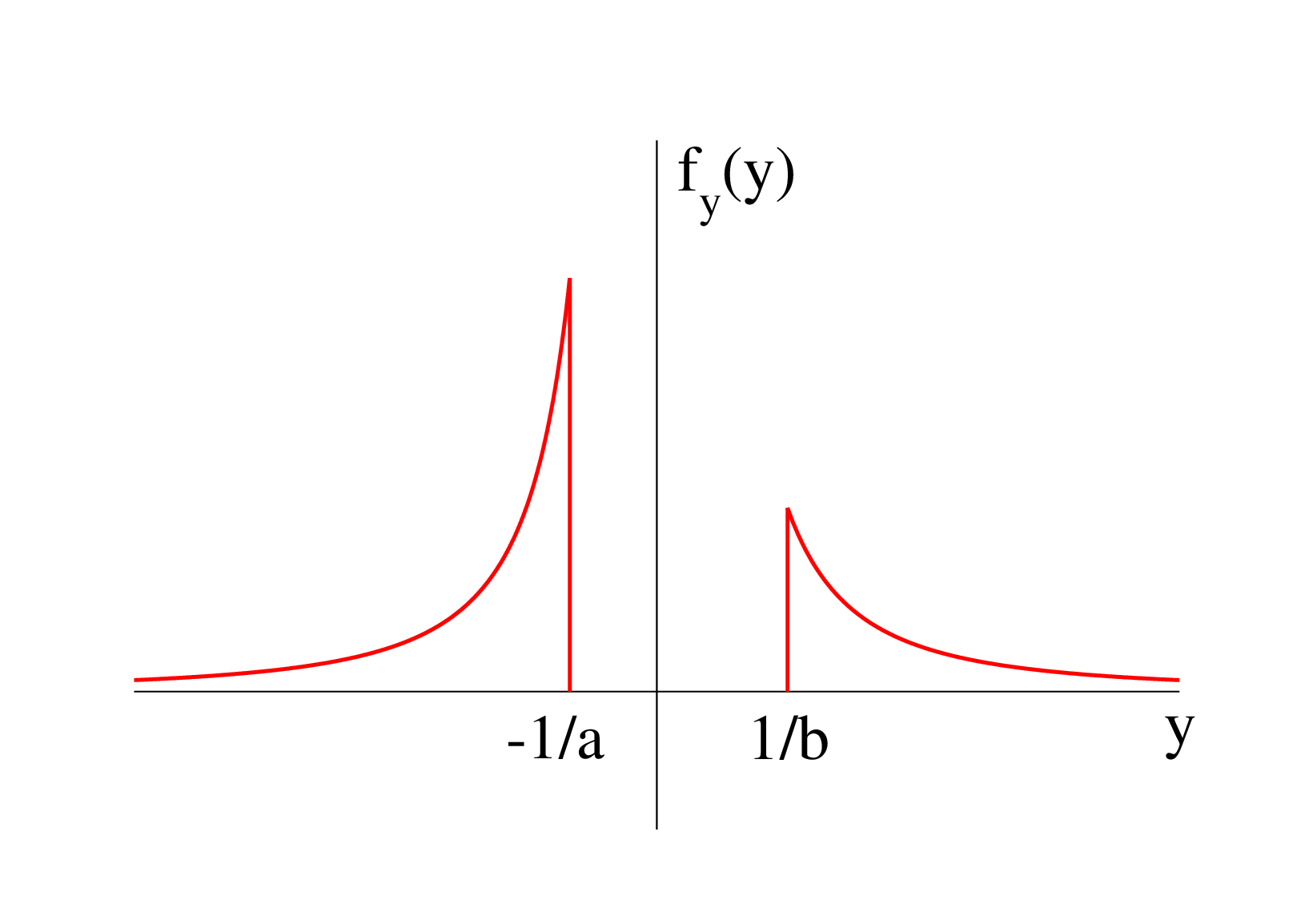}
\caption{\small
Sketchy plot of the probability distributions $f_x(x)$ and $f_y(y)$
given by~(\ref{fxdef}) and~(\ref{fydef}), for $a:b=3:2$.}
\label{rhoxy}
\end{center}
\end{figure}

Let us introduce this notion on a simple and explicit example.
Consider a random variable~$x$ with a uniform distribution
on an interval containing 0, i.e.,
\beq
f_x(x)=\frac{1}{a+b}\qquad(-a<x<b).
\label{fxdef}
\eeq
Its reciprocal $y=1/x$ has the probability distribution
\beq
f_y(y)=\frac{1}{(a+b)y^2}\qquad(y<-1/a\ \ \&\ \ y>1/b).
\label{fydef}
\eeq
These distributions are sketched in figure~\ref{rhoxy}.

The occurrence of symmetric Cauchy tails falling off as $1/y^2$ as $y\to\pm\infty$
is a generic feature of the distribution of a variable $y$ equal to the reciprocal
of a variable~$x$ which may vanish.
The strength $G$ of these Cauchy tails,
or the {\it Cauchy strength} of the variable $y$ for short, is defined as
\beq
G=\lim_{y\to\pm\infty}y^2f_y(y)=\bigmean{\delta\left(\frac{1}{y}\right)}.
\eeq
In the above example, this reads
\beq
G=\mean{\delta(x)}=f_x(0)=\frac{1}{a+b},
\eeq
in agreement with~(\ref{fydef}).

Consider for definiteness the regime where $a$ and $b$ are simultaneously large.
There, $x$ is typically large, and so $y$ is typically small.
There are however rare circumstances where $x$ is close to zero,
and so $y$ is much larger than its typical value.
The Cauchy strength $G$, introduced above, characterizes atypical large values of $y$,
whereas its typical small values are characterized by its symmetric mean value, i.e.,
\beq
H=\mean{y}=\lim_{\Lambda\to\infty}\int_{-\Lambda}^\Lambda y\,f_y(y)\,\dd y.
\eeq
In the above example, we have
\beq
H=\bigmean{\frac{1}{x}}
=\dashint_{-\infty}^\infty\frac{f_x(x)\,\dd x}{x}
=\lim_{\eps\to0}\left(\int_{-\infty}^{-\eps}+\int_\eps^\infty\right)\frac{f_x(x)\,\dd x}{x},
\eeq
where the bar denotes the Cauchy principal value, i.e., explicitly,
\beq
H=\lim_{\eps\to0}\left(\int_{-a}^{-\eps}+\int_\eps^b\right)\frac{\dd x}{(a+b)x}
=\frac{1}{a+b}\ln\frac{b}{a}.
\eeq
This expression vanishes in the symmetric situation where $a=b$, as should be.

In spite of their interpretation,
the Cauchy strength $G$ and the mean value $H$ of $y$ have the same dimension.
As a consequence, they are of the same order of magnitude
in the weak-disorder regime where $a$ and $b$ are simultaneously large.

\section{Effective disorder on the diamond chain and on the pyrochlore ladder}
\label{appeff}

This appendix is devoted to an investigation of
the effective disorder on the diamond chain and the pyrochlore ladder.
We consider a symmetric continuous distribution~$\rho(v)$
of the random potentials, such that all moments are finite.
The disorder width $w$ is defined by setting $\mean{v^2}=w^2$.
The kurtosis is defined by
\beq
\kappa=\frac{\mean{v^4}}{\mean{v^2}^2}=\frac{\mean{v^4}}{w^4}.
\eeq
The Fourier transform of the distribution reads
\beq
\hrho(q)=\mean{\e^{\ii qv}}=\int_{-\infty}^\infty\e^{\ii qv}\rho(v)\dd v.
\eeq

\subsection{The effective disorder on the diamond chain}
\label{appdiam}

The effective disorder on the diamond chain is characterized
by its Cauchy strength and its mean value (see~(\ref{g1h1})):
\beq
G_1(E)=\bigmean{\delta\left(\frac{\eta^A+\eta^B}{\eta^A\eta^B}\right)},
\qquad
H_1(E)=\bigmean{\frac{\eta^A\eta^B}{\eta^A+\eta^B}},
\label{ghdef}
\eeq
with
\beq
\eta^A=E-v^A,\qquad
\eta^B=E-v^B.
\label{etadef}
\eeq

The Cauchy strength can be evaluated as
\beqa
G_1(E)
&=&\int_{-\infty}^\infty\rho(E-\eta^A)\dd\eta^A\int_{-\infty}^\infty\rho(E-\eta^B)\dd\eta^B
\delta\left(\frac{\eta^A+\eta^B}{\eta^A\eta^B}\right)
\nonumber\\
&=&\int_{-\infty}^\infty\rho(E-\eta)\rho(E+\eta)\eta^2\dd\eta.
\label{g1res}
\eeqa
The second line is obtained by integrating over $\eta^B$, denoting $\eta^A$ as $\eta$.
This can be recast in Fourier space as
\beq
G_1(E)=-\frac{1}{2\pi}\int_{-\infty}^\infty\hrho(q)\e^{\ii qE}
\frac{\dd^2}{\dd q^2}\left(\hrho(q)\e^{\ii qE}\right)\dd q.
\label{g1fou}
\eeq
The latter expression is more amenable to an explicit evaluation for specific disorder distributions.

The mean value can be evaluated as
\beq
H_1(E)
=\bigmean{\frac{(E-v^A)(E-v^B)}{2E-v^A-v^B}}
=\frac{E}{2}-\frac{1}{4}\bigmean{\frac{(v^A-v^B)^2}{2E-v^A-v^B}}.
\eeq
At this stage it is advantageous to add an infinitesimal imaginary part to $E$.
We have
\beqa
\Im H_1(E+\ii 0)
&=&\frac{\pi}{4}\bigmean{(v^A-v^B)^2\delta(2E-v^A-v^B)}
\nonumber\\
&=&\frac{\pi}{4}\int_{-\infty}^\infty\rho(v^A)\dd v^A\int_{-\infty}^\infty\rho(v^B)\dd v^B
(v^A-v^B)^2
\nonumber\\
&\times&\delta(2E-v^A-v^B).
\eeqa
Setting $v^A=E-\eta$, $v^B=E+\eta$, we get
\beq
\Im H_1(E+\ii 0)
=\pi\int_{-\infty}^\infty\rho(E-\eta)\rho(E+\eta)\eta^2\dd\eta
=\pi G_1(E).
\eeq
The real part $H_1(E)\equiv\Re H_1(E+\ii 0)$ can be reconstructed from the imaginary part
by an integral equation somehow similar to the Kramers-Kronig dispersion relations:
\beq
H_1(E)=\frac{E}{2}+\dashint_{-\infty}^\infty\frac{G_1(E')\dd E'}{E'-E},
\label{h1res}
\eeq
where the bar again denotes the Cauchy principal value.

The scaling dependence of $G_1(E)$ and $H_1(E)$ on $w$ reads
\beq
G_1(E)=wg_1(x),\qquad H_1(E)=wh_1(x),\qquad x=\frac{E}{w}.
\label{g1h1app}
\eeq
The functions $G_1(E)$ and $g_1(x)$ are even, whereas $H_1(E)$ and $h_1(x)$ are odd,
and~(\ref{h1res}) translates to
\beq
h_1(x)=\frac{x}{2}+\dashint_{-\infty}^\infty\frac{g_1(x')\dd x'}{x'-x}.
\label{h1r}
\eeq

\subsection{The effective disorder on the pyrochlore ladder}
\label{apppyro}

The effective disorder on the pyrochlore ladder
is characterized by its Cauchy strength and its mean value (see~(\ref{g2h2})):
\beq
G_2(E)=\bigmean{\delta\left(\frac{\eta^\1+\eta^\2}{\eta^\1\eta^\2}\right)},
\qquad
H_2(E)=\bigmean{\frac{\eta^\1\eta^\2}{\eta^\1+\eta^\2}},
\label{g2h2def}
\eeq
with
\beq
\eta^\1=2E-v^A-v^C,\qquad
\eta^\2=2E-v^B-v^D.
\label{eta2def}
\eeq
The expressions~(\ref{g2h2def}) and~(\ref{eta2def})
are similar to~(\ref{ghdef}) and~(\ref{etadef}),
up to two differences:
energy is doubled ($E\to2E$),
and each random potential is replaced by a sum of two independent potentials
($v^A\to v^A+v^C$, $v^B\to v^B+v^D$).
The latter operation amounts to replacing the disorder distribution $\rho$
by $\rho_2=\rho*\rho$, i.e.,
\beq
\rho_2(v)=\int_{-\infty}^\infty\rho(u)\rho(v-u)\dd u,
\eeq
whose Fourier transform is $\hrho_2(q)=\hrho(q)^2$.
Applying these changes to the results~(\ref{g1res}), (\ref{g1fou})
and~(\ref{h1res}), we readily obtain
\beqa
G_2(E)
&=&\int_{-\infty}^\infty\rho_2(2E-\eta)\rho_2(2E+\eta)\eta^2\dd\eta
\nonumber\\
&=&-\frac{1}{2\pi}\int_{-\infty}^\infty\hrho(q)^2\e^{2\ii qE}
\frac{\dd^2}{\dd q^2}\left(\hrho(q)^2\e^{2\ii qE}\right)\dd q
\label{g2res}
\eeqa
and
\beq
H_2(E)=E+\dashint_{-\infty}^\infty\frac{G_2(E')\dd E'}{E'-E}.
\label{h2res}
\eeq
The scaling dependence of $G_2(E)$ and $H_2(E)$ on $w$ reads
\beq
G_2(E)=wg_2(x),\qquad H_2(E)=wh_2(x),\qquad x=\frac{E}{w}.
\label{g2h2app}
\eeq
The functions $G_2(E)$ and $g_2(x)$ are even,
whereas $H_2(E)$ and $h_2(x)$ are odd,
and~(\ref{h2res}) translates to
\beq
h_2(x)=x+\dashint_{-\infty}^\infty\frac{g_2(x')\dd x'}{x'-x}.
\label{h2r}
\eeq

\subsection{Explicit formulas for specific disorder distributions}
\label{appdis}

In the body of this work
we use the following results for three disorder distributions:
uniform, Gaussian and exponential.

\subsubsection*{The uniform distribution.}
It reads
\beq
\rho(v)=\frac{1}{2V}\qquad(-V<v<V).
\eeq
We have $\mean{v^2}=V^2/3$ and $\mean{v^4}=V^2/5$, and so
\beq
w=\frac{V}{\sqrt{3}},\qquad\kappa=\frac{9}{5}.
\eeq
Its Fourier transform reads
\beq
\hrho(q)=\frac{\sin qV}{qV}.
\eeq
Inserting the latter expression into~(\ref{g1fou}) and~(\ref{g2res}),
we obtain the following expressions for $g_1(x)$ and $g_2(x)$ for $x\ge0$ (these functions are even):
\beq
g_1(x)=\left\{
\matrix{
\frad{1}{18}(\sqrt{3}-x)^3\hfill &(0<x<\sqrt{3}),\hfill\cr\cr
0\hfill &(x>\sqrt{3}),\hfill
}\right.
\label{unig1}
\eeq
\beq
g_2(x)=\left\{
\matrix{
g_\star(x)\hfill &
\left(0<x<\sqrt{3}/2\right),\hfill\cr\cr
\frad{8}{135}(\sqrt{3}-x)^5\hfill &\left(\sqrt{3}/2<x<\sqrt{3}\right),\hfill\cr\cr
0\hfill &\;(x>\sqrt{3}),\hfill
}\right.
\label{unig2}
\eeq
with
\beq
g_\star(x)=\frac{2\sqrt{3}}{15}-\frac{4\sqrt{3}}{9}x^2+\frac{8}{9}x^3-\frac{8\sqrt{3}}{27}x^4+\frac{8}{45}x^5.
\eeq
The vanishing of~$g_1(x)$ and $g_2(x)$ for $\abs{x}>\sqrt{3}$,
i.e., $\abs{E}>V$, is expected.
Indeed in this range none of the $\eta$ variables
entering~(\ref{etadef}) or~(\ref{eta2def}) can vanish.
The functions $h_1(x)$ and $h_2(x)$ have not been evaluated in closed form.

\subsubsection*{The Gaussian distribution.}
It reads
\beq
\rho(v)=\frac{\e^{-v^2/(2w^2)}}{w\sqrt{2\pi}}.
\eeq
We have $\mean{v^2}=w^2$ and $\mean{v^4}=3w^4$, and so
\beq
\kappa=3.
\eeq
Its Fourier transform reads
\beq
\hrho(q)=\e^{-q^2w^2/2}.
\eeq
Inserting this expression into~(\ref{g1fou}) and~(\ref{g2res}),
we obtain the simple results
\beq
g_1(x)=\frac{\e^{-x^2}}{4\sqrt{\pi}},\qquad
g_2(x)=\frac{\e^{-2x^2}}{\sqrt{8\pi}}.
\eeq
The expressions~(\ref{h1r}) and~(\ref{h2r}) yield
\beqa
h_1(x)&=&\frac{x}{2}+\frac{\ii\sqrt{\pi}}{4}\e^{-x^2}\erf(\ii x),
\nonumber\\
h_2(x)&=&x+\frac{\ii\sqrt{2\pi}}{4}\e^{-2x^2}\erf(\ii x\sqrt{2}),
\eeqa
where $\erf$ is the error function.

\subsubsection*{The exponential distribution.}
The symmetric exponential distribution,
also referred to as the Laplace distribution, reads
\beq
\rho(v)=\frac{\e^{-\abs{v}/V}}{2V}.
\eeq
We have $\mean{v^2}=2V^2$ and $\mean{v^4}=24V^4$, and so
\beq
w=V\sqrt{2},\qquad\kappa=6.
\eeq
Its Fourier transform reads
\beq
\hrho(q)=\frac{1}{1+q^2V^2}.
\eeq
Inserting the latter expression into~(\ref{g1fou}) and~(\ref{g2res}),
we obtain the following expressions for $g_1(x)$ and $g_2(x)$ for $x\ge0$ (these functions are even):
\beqa
g_1(x)&=&
\left(\frac{\sqrt{2}}{16}+\frac{1}{4}x+\frac{\sqrt{2}}{4}x^2+\frac{1}{3}x^3\right)
\e^{-2\sqrt{2}x},
\\
g_2(x)&=&
\left(\frac{7\sqrt{2}}{64}+\frac{7}{8}x+\frac{13\sqrt{2}}{8}x^2+\frac{11}{3}x^3
+\frac{8\sqrt{2}}{3}x^4+\frac{32}{15}x^5\right)
\e^{-4\sqrt{2}x}.
\nonumber
\eeqa
The functions $h_1(x)$ and $h_2(x)$ have not been evaluated in closed form.

\medskip

To close, we give in table~\ref{tabdis} the values of the Cauchy strength
of the effective disorder at the center of the flat band at $E=0$
for the two flat-band structures and the three disorder distributions studied in this appendix.

\begin{table}[!ht]
\begin{center}
\begin{tabular}{|l|c|c|c|}
\hline
Distribution & $\kappa$ & $g_1(0)$ & $g_2(0)$\\
\hline
Uniform & $\frad{9}{5}=1.8$ & $\frad{\sqrt{3}}{6}=0.288675$ & $\frad{2\sqrt{3}}{15}=0.230940$\\
\hline
Gaussian & 3 & $\frad{1}{4\sqrt{\pi}}=0.141047$ & $\frad{1}{\sqrt{8\pi}}=0.199471$\\
\hline
Exponential & 6 & $\frad{\sqrt{2}}{16}=0.088388$ & $\frad{7\sqrt{2}}{64}=0.154679$\\
\hline
\end{tabular}
\caption{Values of the Cauchy strength of the effective disorder at the center of the flat band at $E=0$
for the two flat-band structures studied in this appendix,
namely $g_1(0)$ on the diamond chain (see section~\ref{secdiam} and~\ref{appdiam})
and $g_2(0)$ on the pyrochlore ladder (see section~\ref{secpyro} and~\ref{apppyro}),
and for three disorder distributions,
whose kurtosis $\kappa$ is recalled.}
\label{tabdis}
\end{center}
\end{table}

\section*{References}

\bibliography{paper.bib}

\end{document}